\documentclass[12pt,aps,pra,reprint,groupedaddress,amsmath,amssymb,floatfix,superscriptaddress]{revtex4-2}

\usepackage{graphicx} 
\usepackage{dcolumn} 
\usepackage{bm} 
\usepackage{times}       
\usepackage{mathptmx}    
\usepackage[colorlinks=true, allcolors=blue]{hyperref} 
\usepackage{braket}
\usepackage{qcircuit}
\usepackage[table]{xcolor}
\usepackage[caption=false]{subfig}
\usepackage{enumitem}
\makeatletter
\renewcommand\subsubsection{%
  \@startsection{subsubsection}{3}{0pt}{1.5em}{0.5em}{\bfseries\itshape\centering}}
\makeatother

\begin{document}

\title{Quantum framework for Reinforcement Learning: Integrating  Markov decision process, quantum arithmetic, and trajectory search}

\author{Thet Htar Su}
\email{thethtar@keio.jp, thethtarsu@acsl.ics.keio.ac.jp}
\affiliation{Graduate School of Science and Technology, Keio University, Yokohama, Kanagawa 223-8522, Japan}

\author{Shaswot Shresthamali}
\affiliation{\mbox{Graduate School of Information Science and Electrical Engineering, Kyushu University, Nishi-ku, Fukuoka 819-0395, Japan}}

\author{Masaaki Kondo} 
\affiliation{Graduate School of Science and Technology, Keio University, Yokohama, Kanagawa 223-8522, Japan}
\affiliation{RIKEN Center for Computational Science, Kobe, Hyogo 650-0047, Japan}

\date{December 2024}

\begin{abstract}
This paper introduces a quantum framework for addressing reinforcement learning (RL) tasks, grounded in the quantum principles and leveraging a fully quantum model of the classical Markov decision process (MDP). By employing quantum concepts and a quantum search algorithm, this work presents the implementation and optimization of the agent-environment interactions entirely within the quantum domain, eliminating reliance on classical computations. Key contributions include the quantum-based state transitions, return calculation, and trajectory search mechanism that utilize quantum principles to demonstrate the realization of RL processes through quantum phenomena. The implementation emphasizes the fundamental role of quantum superposition in enhancing computational efficiency for RL tasks. Results demonstrate the capacity of a quantum model to achieve quantum enhancement in RL, highlighting the potential of fully quantum implementations in decision-making tasks. This work not only underscores the applicability of quantum computing in machine learning but also contributes to the field of quantum reinforcement learning (QRL) by offering a robust framework for understanding and exploiting quantum computing in RL systems.
\end{abstract}

\maketitle

\section{Introduction}\vspace{-6pt}
Reinforcement learning (RL) is a branch of machine learning focused on decision-making by autonomous agents ~\cite{sutton2018reinforcementintroduction, graesser2019deepfoundations}. Examples of such agents include robots and self-driving cars. In RL, these agents learn to complete tasks through trial and error without direct human guidance~\cite{Goodfellow-et-al-2016}. By tackling sequential decision-making challenges in uncertain environments, RL holds significant potential for solving complex decision-making problems across various real-world scenarios such as autonomous driving~\cite{shalevshwartz2016}, robotics~\cite{Rl_robotics_a_survey}, and game-playing~\cite{Silver2017, Noam_Superhuman_AI}.  However, classical RL approaches encounter significant challenges in the high-dimensional environments, where the state and action spaces grow exponentially with increasing problem size, making RL computationally expensive~\cite{challenges_RL}. Additionally, training RL models can demand extensive resources and time, particularly when dealing with stochastic environments~\cite{Silver16}. These scalability and computational challenges have motivated the exploration of alternative methods to enhance traditional RL efficiency and effectiveness.

Since advancements in quantum hardware and algorithms offer the capability of integrating quantum computing (QC) with machine learning~\cite{scholten24}, researchers have proposed hybrid quantum-classical methods to address the computational challenges of classical RL. Several studies have introduced quantum-inspired policy optimization algorithms and employed quantum circuits for specific classical components of RL to explore the potential improvements offered by quantum computing in RL~\cite{meyer2022surveyRL}. While hybrid approaches have shown promise, they are often constrained by the additional computational resources required to facilitate communication between classical and quantum systems. This affects the overall performance of the system and limits the full potential of quantum computing. Furthermore, the partial use of quantum techniques leaves the full potential of quantum computing untapped.
These challenges underscore the need for a comprehensive quantum framework for reinforcement learning. We aim to address these bottlenecks associated with classical-quantum interaction and unlock the full potential of quantum mechanics by designing a RL framework where all computations are performed entirely within the quantum domain.

In this work, we propose a complete quantum framework for RL problems, aiming to enhance and extend the principles of classical RL. Classical RL frameworks involve an agent interacting with an environment through a series of states, actions, and rewards, where the agent learns an optimal policy to maximize cumulative rewards over time. This is typically achieved by solving a Markov decision process (MDP), which models state transitions and reward structures. Classical RL often relies on iterative algorithms like Q-learning or policy gradient methods. Building on these foundations, we provide a detailed description of the quantum implementation of a classical MDP, including state transitions that closely mirror those in a classical MDP. Additionally, we describe a quantum approach for calculating cumulative rewards and a trajectory search mechanism leveraging quantum Grover’s algorithm~\cite{groveralgorithm}. The primary contribution of our work lies in demonstrating how reinforcement learning tasks can be solved exclusively using quantum methods, without any classical computations. The contributions of this work are listed as follows:
\begin{enumerate}
[noitemsep]
  \item Quantum representation of MDP: We develop a quantum representation of a classical MDP by applying the principles of quantum superposition. This approach encodes multiple states and actions simultaneously, enabling the system to explore numerous state-action pairs in parallel.
  \item Quantum state transitions: We demonstrate the interactions between the agent and the environment in the quantum domain, illustrating how state transitions are efficiently performed using quantum principles.
  \item Quantum return calculation: We introduce a quantum approach for return calculation that leverages quantum arithmetic. 
  \item Quantum trajectory search: We implement Grover’s algorithm~\cite{groveralgorithm} for trajectory search, facilitating efficient exploration of trajectory sequences in RL tasks. This quantum-enhanced method provides substantial acceleration in identifying optimal trajectories.
\end{enumerate}
This paper is organized as follows: Sec.~\ref{sec:related_work} reviews previous studies on integrating RL with QC. Section~\ref{sec:background} introduces key concepts in classical reinforcement learning and quantum computing. Section~\ref{sec:method} details our quantum implementation, including circuit designs, quantum interactions, and trajectory search. Section~\ref{sec:experiments_results} presents demonstrations and their results, and Sec.~\ref{sec:discussion} and Sec.~\ref{sec:conclusion} summarize insights and propose future research directions.
\vspace{-5pt}

\section{Related Work}\label{sec:related_work}\vspace{-6pt}
Research in quantum reinforcement learning (QRL) covers a variety of approaches, ranging from those that are mainly classical but inspired by quantum principles to those that leverage fully quantum systems. 
\vspace{-10pt}

\subsection{Quantum-inspired reinforcement learning}\vspace{-6pt}
The earliest idea of integrating quantum computing (QC) with reinforcement learning (RL) relied on quantum superposition and amplitude amplification to address the limitations of classical RL, such as slow learning rates, the exploration-exploitation trade-off, and high CPU usage~\cite{meyer2022surveyRL}. Early research focused on representing RL actions as quantum superposition states in value-based algorithms~\cite{Don08,Don06,Chun08,Chen06,Don12}. Subsequently, advancements were made by employing quantum superposition to store policies and utilizing Grover’s algorithm for the action-value function~\cite{Ganger2019,Cho2022}. Recent study has explored the quantum representation of experiences in deep RL~\cite{Wei22}. Researchers have also demonstrated the applicability of quantum-inspired RL (QiRL) approaches in diverse domains, including mobile robotics~\cite{Don06,Don12}, unmanned aerial vehicles (UAVs)~\cite{Li1994}, and clinical decision-making~\cite{Li2020,Niraula2021}. These studies highlight the potential of quantum computing to overcome the limitations of classical RL by achieving a better balance between exploration and exploitation, accelerating the learning process, and reducing computational costs~\cite{meyer2022surveyRL}.
\vspace{-10pt}

\subsection{VQC-based reinforcement learning}\vspace{-6pt}
Many studies have explored combining RL with QC by employing variational quantum circuits (VQCs) as substitutes for classical neural networks in deep RL algorithms to address challenges in large state and action spaces, where the previous QiRL approach encounters limitations~\cite{Sequeira2022}. These efforts were initiated with the use of  VQCs in double deep Q-learning as function approximators for simple discrete environments like Frozen Lake~\cite{Chen20}, and gradually extended to more complex settings, such as Cart Pole~\cite{LockwoodSi20} and Atari games~\cite{lockwood21a}, and continuous action spaces~\cite{Wu_2020}.  Some research~\cite{Skolik2022} has highlighted the importance of architectural design and hyperparameters in shaping RL agent performance which contributed to exponential reductions in model complexity compared to classical methods. Beyond double Q-learning, algorithms like REINFORCE~\cite{Sequeira2022,jerbi_2022} and Actor-Critic~\cite{Kwak21,Lan2021} have successfully employed VQCs to directly approximate policies. These advancements demonstrate that the VQC-based approaches can overcome classical RL limitations by achieving faster convergence, consuming less memory, and utilizing fewer parameters in complex RL environments.
\vspace{-10pt}

\subsection{Reinforcement learning with quantum subroutines}\vspace{-6pt}
Recent advancements in QRL focus on leveraging quantum subroutines on fault-tolerant quantum computers to enhance traditional RL processes. The authors~\cite{wang2021} proposed quantum value iteration algorithm combining quantum mean estimation and quantum maximum finding into a classical value iteration algorithm to reduce sample complexity. Similarly, previous work~\cite{Cherrat2023} introduced quantum policy iteration algorithm, alternating between quantum policy evaluation, which uses quantum linear system solvers to encode the policy value function, and classical policy improvement step based on quantum state measurements. These algorithms demonstrate quadratic speedups in sample complexity compared to classical methods. 
\vspace{-10pt}

\subsection{Full quantum reinforcement learning}\vspace{-6pt}
Recent research~\cite{wiedemann2023} underscores the potential of a fully quantum approach to classical RL through the introduction of a quantum policy iteration framework. This approach evaluates policies by generating a superposition of all possible trajectories within a MDP. By leveraging amplitude estimation for policy evaluation and Grover’s algorithm for policy improvement, the framework significantly reduces the sample complexity of a classical RL algorithm.

In our work, we aim to address the limitations that previous works have encountered. Methods employing VQCs encoded only the agent as a quantum component while retaining the environment in the classical domain, restricting the quantum interactions of agent and environment. Other approaches, such as QiRL and quantum subroutines, relied on classical computations, preventing the realization of fully quantum operations.
Furthermore, while prior work on a fully quantum system~\cite{wiedemann2023} focused on a single-state bandit setting with only one state, our framework extends to more complex decision-making scenarios by implementing a finite-state Markov decision process (MDP) with multiple states, actions, and stochastic transitions. This generalization makes our model applicable to a broader range of practical reinforcement learning problems beyond simple bandit-like tasks. Our framework encodes both the agent and the environment as quantum components, enabling sequential agent-environment interactions across multiple time steps. It incorporates quantum state transitions and return calculations using quantum arithmetic. Our approach enables the quantum agent to explore numerous sequences of interactions simultaneously. Such parallelism enables efficient exploration of trajectories, allowing the agent to evaluate multiple potential outcomes concurrently and accelerate the discovery of high-reward policies.
In contrast to this prior work, which employed Grover’s search to identify high-value actions in a static, single-step scenario, our framework applies Grover’s algorithm to search for an optimal trajectory among all possible full-length quantum trajectories in a multi-state, multi-step MDP. By designing a quantum oracle that marks trajectories based on their cumulative return, our approach enables direct optimization of agent's performance over multiple time steps. This significantly expands the applicability of Grover’s search beyond action-level selection, enabling efficient discovery of optimal policies in multi-step reinforcement learning tasks.

To further advance fully quantum RL, our framework enables direct interaction between the quantum agent and the environment entirely within the quantum domain. The concept of a quantum-accessible environment was first introduced in~\cite{PhysRevLett.117.130501} where the agent and environment interact through an intermediate register. In comparison, our method enables direct communication between the agent and environment, eliminating the need for an additional layer. This direct interaction allows for the simultaneous evaluation of multiple states and actions, increasing computational parallelism and enabling faster exploration of the state-action space. Furthermore, since these interactions are entirely within the quantum domain, this will remove the classical-quantum conversions and achieve the complete quantum interactions. Our framework facilitates optimization by a quantum search algorithm directly to the quantum states generated through these interactions. This approach eliminates reliance on classical subroutines and achieve significant computational speedups. To the best of our knowledge, our work is the first attempt to implement a complete quantum framework for reinforcement learning by integrating the quantum realization of the classical Markov decision process (MDP). 
\vspace{-4pt}

\section{Background}\label{sec:background}\vspace{-6pt}
\subsection{Reinforcement learning}\vspace{-6pt}
Reinforcement learning (RL) is an interaction-based machine learning paradigm, focusing on agents to make sequential decisions in dynamic environments with the goal of maximizing long-term rewards. This paradigm involves an iterative learning process where the agent explores various actions to understand their impact and exploits this knowledge to achieve optimal outcomes~\cite{Plaat_2022}. In RL, an agent interacts with an environment that provides feedback in the form of states and rewards. States reflect the current configuration of the environment, actions represent the agent's possible choices, and rewards are numerical signals received as feedback for the agent's actions, guiding its learning process~\cite{morales2020grokking}.

\begin{figure}[h]
\includegraphics[width=0.45\textwidth]{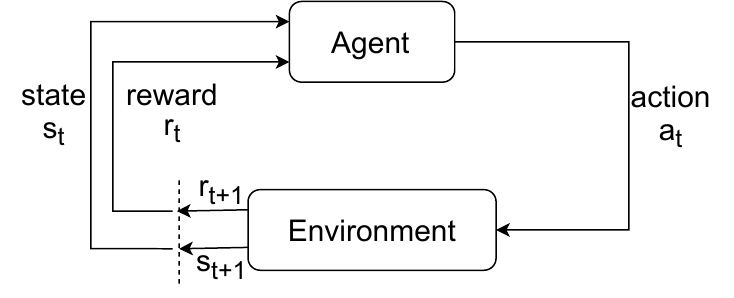}
\caption{The agent-environment interaction in a Markov decision process (MDP)~\cite{sutton2018reinforcementintroduction}.}
\label{fig:mdp}
\end{figure}

RL problems are commonly modeled as Markov decision processes (MDPs), a mathematical framework well-suited for decision-making tasks in uncertain environments. A MDP is formally defined as a 4-tuple \textit{S, A, R(.), P(.)}~\cite{graesser2019deepfoundations}, where \textit{S} is the set of states where the agent can observe from the environment. \textit{A} is the set of actions the agent can execute in the environment. \textit{R(s,a)} is the reward function of the environment. The value $\textit{R(s,a)}:=\mathbb{E}[r_t| s_t=s, a_t=a]$ represents the reward $r_{t}$ received when action $a_{t}$ is performed in state $s_{t}$. Finally, $P(s_{t+1}|s_t,a_t)$ is the state transition function. The value $P(s_{t+1})$ gives the probability that the environment transitions to state $s_{t+1}$, if the agent executes action $a_{t}$  in state $s_t$ at time \textit{t} \cite{meyer2022surveyRL}. Specifically, the agent and environment interact at each of a sequence of discrete time steps, \textit{t = 0, 1, 2, 3, . . }.  At each time step \textit{t}, the agent receives a representation of the environment's state, $s_t \in S$, and based on this, selects an action, $a_t \in A$. The agent receives a numerical reward, $r_t \in R \subset \mathbb{R}$, and transitions to a new state, $s_{t+1} \in S$. The interaction between the agent and the environment generates a trajectory $s_0,a_0,r_0,s_1,a_1,r_1,s_2,a_2,r_2,...$ \cite{sutton2018reinforcementintroduction}. This process is described in Fig.~\ref{fig:mdp}. This process can terminate by reaching a terminal state or a maximum time step $t = T$. 

The primary goal of RL is for the agent to learn optimal actions for each state that optimally maximizes the expected cumulative reward through the interactions with the environment. Balancing exploration (trying new actions) and exploitation (using known actions) is crucial for effective learning in RL, making it a comprehensive and dynamic approach distinct from other machine learning paradigms.

\subsubsection{Q-learning}
RL methods can generally be classified into two broad categories: value-based and policy gradient-based. Value-based methods focus on estimating a value function and derive policies from it. Examples include Q-learning and SARSA (State-Action-Reward-State-Action). Policy-gradient methods directly optimize the policy by maximizing the expected cumulative reward through gradient ascent. Examples include REINFORCE and Actor-Critic methods. In this work, we focus on value-based Q-learning algorithm.  Q-learning is a widely used, model-free reinforcement learning algorithm that estimates the optimal policy by updating a Q-value function. 
The updates are based on the Bellman equation, Eq.~(\ref{eq:background_one}) \cite{sutton2018reinforcementintroduction}.
\begin{equation}
    Q(s, a) =  Q(s, a) + \alpha \left[ r + \gamma \max_{a'} Q(s', a') - Q(s, a) \right]
\label{eq:background_one}.
\end{equation}
The Q-learning equation updates the Q-value $Q(s, a)$, which represents the expected cumulative reward for taking action $a$ in state $s$. $r$ is the immediate reward, and $\gamma$ is the discount factor that weighs future rewards. The term $\displaystyle \max_{a'} Q(s', a') $ represents the maximum Q-value for the next state $s'$ across all possible actions $a'$ in that state. The learning rate $\alpha$ controls how much new information updates the current Q-value. This formulation enables the agent to improve its policy by iteratively refining its value estimates.
\vspace{-10pt}

\subsection{Quantum computing}
\begin{figure}[h]
\includegraphics[width=0.5\textwidth]{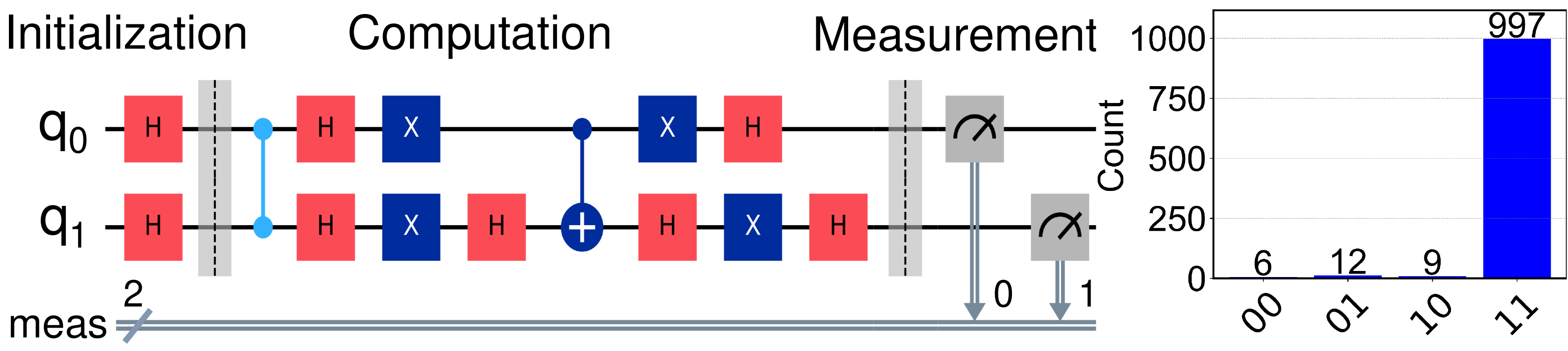}
\caption{Quantum circuit for Grover's algorithm on 2 qubits, searching for the target state $\ket{11}$ generated by Qiskit \cite{qiskit2024}. The measurement was performed using IBM quantum processor (ibm\_brisbane, version: 1.1.62, processor type: Eagle r3, qubits: 127) . Output distribution is displayed on the right, showing the search state $\ket{11}$ with the highest count.}
\label{fig:grover}
\end{figure}

\vspace{-6pt}
Quantum computing (QC) combines quantum mechanics with computer science to enable advanced computation and information processing~\cite{quantum_gentleintro}. In QC, information is represented using qubits, which can exist in superposition states. This unique property allows qubits to process vast amounts of information simultaneously. The quantum computation process consists of three primary steps. The first step is state initialization, where qubits are typically prepared into uniform superposition. The second step involves the application of quantum gates to perform the computation. These gates are represented by unitary matrices, ensuring that transformations applied to quantum states are reversible and that the total probability is preserved~\cite{nielsen2010quantum}.  The final step is measurement, where qubits collapse from their quantum superposition into classical states, producing probabilistic outcomes that represent the results of the computation. These steps collectively enable the construction of complex quantum algorithms capable of solving advanced problems such as Shor’s algorithm for integer factorization~\cite{Shor1994Algorithms,Shor1994Algorithms_97,Shor_1996} and Grover’s algorithm for searching unsorted datasets~\cite{groveralgorithm}. 
In this work, Grover’s algorithm is used as a quantum subroutine to search through possible trajectories and identify those that yield the highest return. Fig.~\ref{fig:grover} presents a simple example of a Grover’s algorithm quantum circuit for 2 qubits, which searches for the target state $\ket{11}$. This circuit illustrates the key steps of the algorithm, which include initialization, computation (oracle marking and amplitude amplification), and measurement. It conceptually demonstrates how Grover’s mechanism drives our quantum trajectory search process presented in Sec.~\ref{sec:experiments_results}.
\vspace{3pt}
\vspace{-28pt}

\section{Method}\label{sec:method}\vspace{-6pt}
\subsection{Quantum implementation of Markov decision process}\vspace{-6pt}
In reinforcement learning, the agent interacts with the environment within a Markov decision process (MDP) framework. The agent selects actions based on a policy, which maps states to actions, while the environment determines state transitions and provides rewards. This interaction is cyclic: the agent observes a state, takes an action, receives a reward, and updates its policy to maximize cumulative rewards over time. Key MDP components include states, actions, a state transition function, and a reward function. For the MDP framework, we consider a simple stochastic environment with four states $(s_0,s_1,s_2,s_3)$ and two actions $(a_0,a_1)$, where the agent moves from one state to another with certain probabilities depending on the chosen action. For example, if the agent is in state $s_0$ and selects action $a_1$, there is some probability that it will move to $s_1$, and some probability it will remain in $s_0$ or will move to another state. This probability of transition from one state to another is determined by the state transition function. The reward function assigns a numerical value to each transition, representing the benefit of moving from one state to another. For instance, transitioning to state $s_1$ might yield a reward $r_1$. The goal of the agent is to find the optimal policy, a strategy that determines which action to take in each state. The optimal policy aims to maximize the total reward the agent can collect over time. Classical Q-learning solves this problem by allowing the agent to learn from its interactions with the environment. The agent explores the states, takes actions, and observes the resulting rewards and transitions. It builds a table, known as a Q-table, that estimates the expected cumulative reward for each state-action pair. By balancing exploration (trying new actions) and exploitation (choosing actions with the highest known rewards), the agent identifies the optimal actions for each state. 

In a classical system, these MDP components are encoded using classical bits. States \textit{S} and actions \textit{A} are elements of finite sets, ranging from simple discrete values to complex vectors. The state transition function is usually expressed as a matrix and rewards are typically regarded as real numbers. 
However, in the quantum system, the MDP formulation lies in the use of the quantum superposition principle and quantum operations, such as unitary matrices, to define the agent, environment, and their interactions. In the following subsection, we first present the initialization of states and actions in the quantum domain. Then, we describe the quantum state transition function and quantum reward function.

\subsubsection{Agent and environment on quantum computer}
\paragraph{State and action initialization.} We are focusing on a classical MDP circuit comprising four states $S = \{s_0, s_1, s_2,s_3\}$ and two actions $A = \{a_0, a_1\}$. In the quantum implementation of this MDP, we define a state space \textit{S} and an action space \textit{A}. For the state space \textit{S}, each state $s \in S $ is encoded as an orthonormal basis vector in a Hilbert space and expressed as a quantum state $\ket{s}$. Similarly, all the possible actions are represented as vectors $\ket{a} \in A$ in the action space \textit{A}. To represent \textit{N} distinct states in a quantum system, \textit{n} qubits are required, where $n= \log_2(N)$, to encode all the possible states into quantum states. Therefore, our quantum MDP necessitates two qubits for the state set \textit{S} and one qubit for the action set \textit{A}. These qubits can be described as strings of qubits on a quantum computer, where each string represents the binary representation of an integer corresponding to that quantum state, i.e.,
\[s_0 = \ket{00} ,s_1 = \ket{01} , s_2 = \ket{10}, s_3 = \ket{11},\] 
\vspace{-20pt}
\[a_0  = \ket{0} ,a_1 = \ket{1}.\]

As we are implementing MDP in the quantum domain, these states and actions can be prepared into superposition by using Hadamard transformation. This process can initialize qubits as a uniform combination of quantum states for all states and actions within their respective spaces. 
For the state space \textit{S}, we define the initial ground state of each state register as $\ket{0_s} = \ket{0}$. Applying the Hadamard transformation to these registers generates a uniform superposition of quantum states, as formulated in Eq.~(\ref{eq:method_one}).
\begin{equation}
   H(\ket{0_s} \otimes \ket{0_s}) = \sum_{n = 0}^{N-1} c_n \ket{s_n}
\label{eq:method_one}.
\end{equation}
The output quantum state is a uniform superposition of the basis states, $\ket{s_0}, \ket{s_1},..., \ket{s_n}$,  within its state space \textit{S}. \textit{N} represents the total number of states in that space. 
Each coefficient $c_n$ represents the amplitude of the corresponding basis state and is uniformly set to $\frac{1}{\sqrt{N}}$, ensuring that the resulting quantum state is properly normalized. This satisfies the normalization condition $\sum_{n=0}^{N-1} \vert c_n \vert ^2 = 1$.  In the case of a 2-qubit state space where $N = 4$, each amplitude becomes $c_n = \frac{1}{2}$.

Similarly, the set of possible actions is encoded using the Hadamard operator to create a uniform superposition of all actions that belong to the action space \textit{A}, as formalized in Eq.~(\ref{eq:method_two}). The initial ground state of the action register is defined as $\ket{0_a}= \ket{0}$, and applying the Hadamard transformation results in a uniform superposition where each amplitude is $c_n = \frac{1}{\sqrt{2}}$, satisfying the normalization condition within the action space.
\begin{equation}
    H\ket{0_a} = \sum_{n=0}^{N-1} c_n \ket{a_n}
    \label{eq:method_two}.
\end{equation}

At the initial stage of the quantum Markov decision process (QMDP), all state and action qubits are initialized into uniform superposition within their respective quantum spaces. This process is described by Eq.~(\ref{eq:method_three}) for the state space \textit{S} and Eq.~(\ref{eq:method_four}) for the action space \textit{A}. After this initialization, the resulting quantum state represents the probability distribution of all possible state-action pairs. Since the states and actions are initialized in uniform superposition, the resulting distribution is also uniform.
\begin{equation}
H(\ket{0_s} \otimes \ket{0_s}) = \frac{1}{2} \ket{00} + \frac{1}{2} \ket{01} + \frac{1}{2} \ket{10} + \frac{1}{2} \ket{11}
\label{eq:method_three}.
\end{equation}
\begin{equation}
    H \ket{0_a} = \frac{1}{\sqrt{2}} \ket{0} + \frac{1}{\sqrt{2}} \ket{1}
\label{eq:method_four}.
\end{equation}

\paragraph{State transition function.} After preparing all states in the state space \textit{S} and all actions in the action space \textit{A} as uniform superposition quantum states, we then consider the state transition function, which defines the likelihood of transitioning from one state to another given a specific action. In a classical MDP, the state transition function $P(s'|s,a)$ determines the probability of transitioning to next state ${s'}$ given the current state $s$ and action $a$. In a quantum circuit, these probabilities are encoded in the amplitudes of quantum states, where the squared magnitudes of the amplitudes correspond to probabilities.
This is achieved by applying $R_y(\theta)$ to an ancillary qubit initialized in the $\ket{0}$ state, producing a superposition in which the probabilities of measuring $\ket{0}$ and $\ket{1}$ are $\cos^2\left(\frac{\theta}{2}\right) \text{ and } \sin^2\left(\frac{\theta}{2}\right)$, respectively. 
To encode classical transition probabilities $P(s'|s,a)$, the rotation angle $\theta$ is calculated as $\theta = 2\arcsin(\sqrt{P(s' | s, a)})$. In our work, we use multi-controlled $R_y(\theta)$ gates to apply these rotations conditionally based on the current state $\ket{s}$ and action $\ket{a}$. The rotation is applied only when the control qubits match a specific state-action pair $(\ket{s^*},\ket{a^*})$, where $\ket{s^*} \in \textit{S}$, $\ket{a^*} \in \textit{A}$. Equation~(\ref{eq:method_six}) is a mathematical representation of this quantum state transition process where the controlled $R_y(\theta)$ gates apply a rotation to an ancillary qubit $\ket{0_{s'}}$ by an angle $\theta$, depending on the state-action pair, thereby determining the next state, $\ket{s'} \in S$. 
\begin{equation}
    CR_y(\theta) (\ket{s} \ket{a} \ket{0_{s'}}) =
    \begin{cases}
        \ket{s} \ket{a} R_y(\theta) \ket{0_{s'}} & \text{if} \ket{s}=\ket{s^*},\\ & \hspace{7pt} \ket{a}=\ket{a^*}, \\   
        \ket{s} \ket{a} \ket{0_{s'}}  & \text{else}.
    \end{cases}
\label{eq:method_six}
\end{equation}

\paragraph{Reward function.} 
The reward function in a classical MDP is defined as $R(s,a)$, where specific state-action pairs $(s,a)$ determine a reward value. The Controlled-NOT (CNOT) gate models this behavior conditionally, where the control qubits (representing the current state $\ket{s}$ and action $\ket{a}$) dictate whether the target qubit (representing the reward $\ket{r}$ ) is flipped. In our quantum circuit, the next state qubits $\ket{s'}$ are used as control qubits since they represent the result of transitioning from a specific state-action pair to a next state $\ket{s'}$, implicitly encoding the state-action pair $(s,a)$ in determining the reward. The CNOT gate works conditionally such that the reward qubit $\ket{r}$, initially set to $\ket{0}$, is flipped to $\ket{1}$ if the next state matches a condition where a reward exists. This ensures that the reward qubit reflects the reward function based on the transition outcome (next state), determined by the state-action pair. The mathematical formulation of the reward function can be expressed as Eq.~(\ref{eq:method_seven}) where the CNOT gate flips an ancillary qubit $\ket{0_r}$ when the control qubits match valid next states $\ket{s'}$, thereby generating the corresponding reward.
\begin{equation}
        CNOT(\ket{s'} \ket{0_r}) = \ket{s'} \ket{s'\oplus 0_r}           
\label{eq:method_seven}
\end{equation}
\vspace{-15pt}

\subsubsection{Agent and environment single interaction on quantum computer}
Building upon the quantum implementation of MDP components discussed earlier, the interaction between the agent and the environment in a quantum Markov decision process (QMDP) is governed by quantum dynamics and can be understood as a sequence of quantum operations. 

At each interaction step, the agent receives the current state $\ket{s}$ of the environment. Then, the agent applies a unitary operation that corresponds to its policy $\pi$ encoded as $\ket{a}$. This process of action selection influences the evolution of the environment’s state, resulting in a transition of the system to the next state $\ket{s'}$. Then, the agent receives the reward $\ket{r}$ based on the next state $\ket{s'}$. In the context of QMDP, this agent-environment interaction can be modeled using a unitary operator $U(S\otimes A\otimes S\otimes R)$, which prepares a quantum state representing the distribution of trajectories for the state, action, next state and reward sequence. This quantum state can be expressed as follows in Eq.~(\ref{eq:method_nine}).
\begin{equation}
        \ket{\phi} = \sum_{n=1}^N \bigg[c_{s',r|s,a} \ket{s} \ket{a}\ket{s'}\ket{r} \bigg]^n
\label{eq:method_nine}.
\end{equation} 
where $c_{s',r|s,a}$ represents the amplitude that the system transitions to the next state $\ket{s'}$ with reward $\ket{r}$, given by the current state $\ket{s}$ and action $\ket{a}$. The square of the amplitude $\vert c_{s',r|s,a} \vert^2$ corresponds to the probability of the transition and $N$ represents the total number of trajectories in that quantum interaction.
\vspace{-5pt}

\subsection{Agent-environment quantum interactions over multiple time steps}\vspace{-6pt}
To extend the QMDP framework over $\textit{T}$ time steps, we need to account for the quantum state transition and the calculation of cumulative rewards. At each time step $\textit{t}$, the agent receives a state $\ket{s_t}$, performs an action $\ket{a_t}$, and the environment responds by transitioning to a new state $\ket{s'_t}$ while providing a reward $\ket{r_t}$. This process is repeated over $\textit{T}$ time steps, ensuring that the next state $\ket{s'_t}$ at $\textit{t}$ becomes the current state $\ket{s_t}$ for the subsequent time step.  

In a quantum circuit, this state transition can be seen as transferring the outcome of the next state qubits (representing $\ket{s'_t}$) to the current state qubits for the next time step (denoting $\ket{s_{t+1}}$). The CNOT gate can be used to propagate these states because it conditionally transfers information in a quantum system. It works well for state transitions because if the control qubit ($\ket{s'_t}$) is $\ket{0}$, the target qubit ($\ket{s_{t+1}}$) remains unchanged; however, if the control qubit ($\ket{s'_t}$) is $\ket{1}$, the target qubit ($\ket{s_{t+1}}$) is flipped. This effectively copies the state of $\ket{s'_t}$ onto $\ket{s_{t+1}}$ in a way that the next state at $t=0$ becomes the current state at $t=1$. This behavior ensures that information about the next state $\ket{s'_t}$ is conditionally propagated to the target qubits (current state qubit $\ket{s_{t+1}}$) for the next time step while preserving the state information. In our quantum circuit, where states, actions, and next states exist in superposition, the CNOT gate can also preserve this quantum superposition and enable all possible state transitions to occur simultaneously.

Below is a description of how this process unfolds over $T$ time steps, including the corresponding quantum operators that govern state transitions and rewards. The approach builds on the single interaction discussed earlier and extends it to a sequence of interactions.
Initially at time step ${t=0}$, the agent starts at state $\ket{s_0}$, takes an action $\ket{a_0}$, transition to the next state $\ket{s'_0}$ and receive a reward $\ket{r_0}$. The next state $\ket{s'_0}$ propagates to the subsequent time step. At time step ${t=1}$, the agent observes the new state $\ket{s_1}$ generated at ${t=0}$, selects a new action $\ket{a_1}$, and the environment transitions to the next state $\ket{s'_1}$ and provides the reward $\ket{r_1}$. Once again, the next state $\ket{s'_1}$ becomes the current state for the subsequent time step ${t=2}$. In this way, the interaction process will occur for $T$ time steps. The quantum state distribution at ${t=1}$ can be expressed as in Eq.~(\ref{eq:method_eleven}), where $N$ represents the total number of trajectories in that quantum interaction. The terms $c_{s'_0,r_0 | s_0,a_0}$ and $c_{s'_1,r_1| s_1, a_1}$ are the amplitudes indicating that the system transitions to the next state $\ket{s'}$ with reward $\ket{r}$, given current state $\ket{s}$ and action $\ket{a}$ at time step ${t=0}$ and ${t=1}$, respectively.
\begin{align}
    \ket{\phi^{(1)}} &= \sum_{n=1}^{N} \bigg[ \big(c_{s'_0,r_0 | s_0,a_0} + c_{s'_1,r_1| s_1, a_1} \big) \nonumber \\
    &\quad \times \big(\ket{s_0} \ket{a_0} \ket{s'_0}\ket{r_0} \ket{s_1} \ket{a_1}\ket{s'_1}\ket{r_1} \big)\bigg]^n .
\label{eq:method_eleven}
\end{align}

For the $T$ time steps of interaction, each quantum state can be represented as a string of qubits encoding the possible transitions from ${t=0}$ to ${t=T-1}$, formed by concatenating state-action-next state-reward sequences. This is obtained by applying a sequence of individual interactions. The complete quantum distribution of the agent-environment interaction at time step $T$ can be summarized as Eq.~(\ref{eq:method_twelve}).
\begin{equation}
        \ket{\phi^{(T)}} = \sum_{n=1}^N \bigg[\sum_{t=0}^{T-1} c_{s'_t,r_t|s_t,a_t} \prod_{t=0}^{T-1} \ket{s_t} \ket{a_t}\ket{s'_t}\ket{r_t}\bigg]^n
\label{eq:method_twelve}.
\end{equation} 

The entire quantum state after $T$ time steps is a quantum superposition over all possible trajectories $N$, where each trajectory consists of state, action, next state and reward sequences from ${t=0}$ to ${t=T-1}$. The amplitudes $\sum_{t=0}^{T-1} c_{s'_t,r_t|s_t,a_t}$ define the likelihood of each trajectory based on the agent's actions and environment's responses up to time step $T$. 

\subsubsection{Return calculation by quantum arithmetic}
In classical reinforcement learning (RL), the reward is a key signal used to guide the agent's behavior. It is a scalar value that the agent receives from the environment as feedback for taking an action in each state. The goal of the agent is to maximize the cumulative reward over time. This total accumulated reward the agent receives over a trajectory of interactions is called a return. This return is typically discounted by a factor $\gamma\in [0,1]$, which prioritizes immediate rewards over future rewards. 

In our work, this classical concept is mapped into the quantum domain, where the return is encoded as a quantum state $\ket{g}$  in a Hilbert space $G$. This space represents a quantum system with a sufficient number of qubits to encode every potential return value. Initially, the quantum state $\ket{g}$ is prepared in its ground state $\ket{0_g}$. To compute return, the quantum operation $U_G$ is applied to the reward qubits $\ket{r_t}$ for each time step. This operation executes a quantum addition that performs sequential bit-wise addition of each reward register $\ket{r_t}$ and stores the result in the return registers $\ket{g}$. This operation propagates carry bits through the circuit, similar to a classical addition. For each reward register $\ket{r_t}$, a series of CNOT and Toffoli gates are applied to implement bit-wise addition into the return registers. Conceptually, this operation accumulates the discounted sum of rewards across T time steps. 
An abstract representation of this transformation is given in Eq.~(\ref{eq:method_thirteen}), where $\gamma\in [0,1]$  is a fixed discount factor, $T$  is the total number of time steps, and $\ket{r_t}$  represents the reward at time step $t$.
\begin{equation}
    U_G := \sum_{t=0}^{T-1} \gamma \ket{g} \bra{r_t}
\label{eq:method_thirteen}.
\end{equation}
This equation follows the standard form of discounted return in classical RL, but here, the rewards $\ket{r_t}$ are encoded in quantum registers as qubits and the sum of discounted rewards is then calculated using quantum arithmetic. This summation result for each trajectory is encapsulated in the quantum register $\ket{g}$.

When applying the return operation $U_G$ to the quantum Markov decision process (QMDP) over $T$ time steps, the goal is to compute the return for each possible trajectory by summing the discounted rewards across those time steps.
The final quantum state over these interactions can be expressed as Eq.~(\ref{eq:method_fourteen}).
\begin{equation}
        \ket{\phi^{(T)}} = \sum_{n=1}^N \bigg[\bigg[\sum_{t=0}^{T-1} c_{s'_t,r_t|s_t,a_t} \prod_{t=0}^{T-1} \ket{s_t} \ket{a_t}\ket{s'_t}\ket{r_t}\bigg] \otimes \ket{g}\bigg]^n
\label{eq:method_fourteen}.
\end{equation} 
where N is the total number of possible quantum trajectories for the agent-environment interactions over $T$ time steps, $\ket{g}$ represents the return for each trajectory, which sums the discounted rewards over the $T$ time steps. Therefore, the complete quantum state $\ket{\phi^{(T)}}$ contains not only the state, action, next state and reward sequences but also the return for each trajectory. This enables the QMDP to evaluate the performance of all possible sequences of interactions efficiently by computing the return across multiple trajectories simultaneously.

The quantum state $\ket{\phi^{(T)}}$, discussed in Eq.~(\ref{eq:method_fourteen}), is not human-interpretable classical information. The coefficients $c_{s'_t,r_t|s_t,a_t}$ that define the amplitude of the quantum state cannot be accessed directly due to the nature of quantum systems. The only way to obtain the information from the quantum state $\ket{\phi^{(T)}}$ is by measurement. Measurement is a probabilistic process that results a specific quantum state with a likelihood determined by $\vert c_{s'_t,r_t|s_t,a_t}\vert ^2$ . Therefore, the quantum state $\ket{\phi^{(T)}}$ can be regarded as quantum encoding of probability distributions over the possible trajectories of the agent-environment interactions across $T$ time steps.
\vspace{-5pt}

\subsection{Quantum trajectories search}\vspace{-6pt}
In the context of classical RL, the primary objective is to find the optimal policy that dictates the best action in each state to maximize the long-term return. Classical methods for finding optimal policies include dynamic programming techniques such as value iteration and policy iteration, and model-free methods like Q-learning and SARSA~\cite{sutton2018reinforcementintroduction}.

In quantum reinforcement learning (QRL), we apply the power of quantum algorithms to efficiently search for desired trajectories and identify the best action for each state. We utilize Grover's search~\cite{groveralgorithm} to identify optimal quantum trajectories from a set of agent-environment interactions over multiple time steps. Each trajectory consists of the agent’s action, the resulting quantum state of the environment, and a return. The algorithm seeks to find optimal quantum trajectories, which can inform the optimal policies for a quantum reinforcement learning (QRL) agent. 

We formulated the interaction between a quantum agent and its environment over $T$ time steps (Eq.~(\ref{eq:method_fourteen})). A quantum trajectory here refers to the series of states and actions the agent takes, together with the environment's responses. Each trajectory $\ket{T}$ can be represented as $\ket{T} = \ket{s_0,a_0,s'_0,r_0,s_1,a_1,s'_1,r_1, \ldots, s_{T-1},a_{T-1},s'_{T-1},r_{T-1},g}$ where $s$ is the current state at each time step from $t=0$ to $t=T-1$, $a$ is the action taken by the agent, $s'$ is the next state reached after taking an action at state $s$, $r$ is the reward received and $g$ is the total return throughout the trajectory. 

Our goal is to find the optimal trajectory from this dataset, which maximizes the return $\ket{g}$.  To efficiently identify the optimal quantum trajectory, we implement Grover's search algorithm~\cite{groveralgorithm}. Firstly, we define an oracle whose primary purpose is to identify and mark the desired solutions within this search dataset. In our case, this involves marking quantum trajectories that are likely to receive maximum return. The oracle can be represented as a unitary operator $U_w$, which flips the phase of the desired trajectory. Mathematically, this operation can be described as in Eq.~(\ref{eq:method_sixteen}), where $\ket{T}$ represents a quantum trajectory and $\ket{\psi}$ is the target quantum state. 
\begin{equation}
  U_w \ket{T} = 
  \begin{cases}
        -\ket{T}  & \text{if $\ket{T}$ = $\ket{\psi}$},\\
          \ket{T}  & \text{otherwise}. 
    \end{cases}
\label{eq:method_sixteen}
\end{equation}

Once the oracle has marked the desired trajectories, the next step involves amplifying the amplitude of these marked states. This step is done by quantum amplitude amplification~\cite{quantum_intro} which increases the probability of measuring the desired solutions. The process of amplification $U_s$ can be expressed as Eq.~(\ref{eq:method_seventeen}) \cite{Guo_2023}:
\begin{equation}
  U_s = 2 \ket{\psi_s} \bra{\psi_s} - I
\label{eq:method_seventeen}.
\end{equation}
where $\ket{\psi_s}$ represents the superposition of overall quantum trajectories, and $I$ is the identity operator. Applying Grover's algorithm, which involves repeating the oracle $U_w$ and amplification $U_s$ operations, the probability of measuring the desired trajectory is maximized after a certain number of iterations.
\vspace{-5pt}

\section{Demonstrations and Results}\label{sec:experiments_results}
\begin{figure}[h]
    \includegraphics[width=0.45\textwidth]{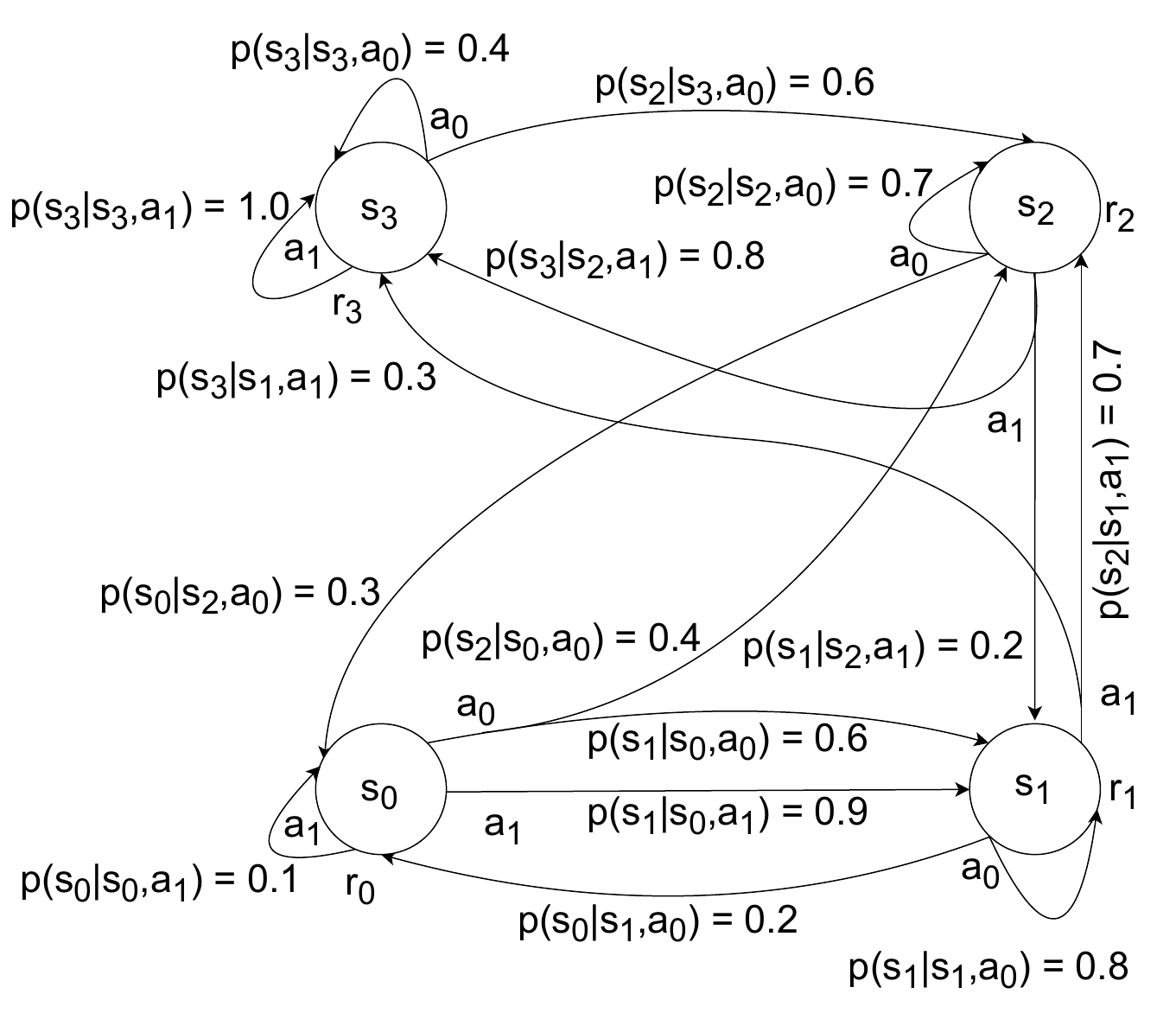}\vspace{-5pt}
    \caption{Graphical representation of a classical MDP with four states $(s_0,s_1,s_2,s_3)$, two actions $(a_0, a_1)$ and rewards $(r_0, r_1, r_2, r_3)$. The arrows between states indicate the transitions associated with each action. Transition probabilities are labeled on each arrow.}
    \label{fig:state_transition}
\end{figure}

\vspace{-6pt}
In the demonstrations and results section of this work, we first present the dynamics of the classical Markov decision process (MDP) through a graphical representation of state transitions and associated probabilities, as shown in Fig.~\ref{fig:state_transition}. The diagram includes four states labeled $s_0,s_1,s_2$ and $s_3$, where $s_3$ represents the terminal state, and two actions, $a_0$ and $a_1$. The arrows connecting the states indicate the transitions between these states for a given action. Each transition is labeled with a probability value that denotes the likelihood of moving from one state to another under a specific action. These probabilities provide insights into the stochastic nature of the state transitions. For example, transitioning from $s_0$ to $s_1$ with action $a_0$ has a probability of 0.6. Similarly, transitioning from $s_0$ to $s_2$ under action $a_0$ occurs with a probability of 0.4. The diagram also illustrates several self-transitions, such as the one in state $s_3$, where action $a_1$ leads to a transition back to $s_3$ with a probability of 1. These self-transitions capture situations where the system remains in the same state regardless of the action taken. The figure visualizes the probabilistic behavior of the system under different actions, as well as how the state evolves over time based on these actions.
\vspace{-5pt}

This classical MDP serves as a reference framework, laying the groundwork for comparing it to the quantum version in our proposed QMDP model. 
The quantum model introduces superposition effects, enhancing the decision-making process with additional computational capabilities. Specifically, the superposition principle allows the quantum model to represent and process all possible state-action combinations in parallel. This enables the agent to explore multiple trajectories simultaneously, significantly reducing the number of interactions required to learn optimal policies.  The quantum implementation further benefits from Grover’s search algorithm, which provides a speedup for identifying optimal trajectories based on cumulative returns. These computational advantages go beyond classical capabilities by allowing efficient encoding, evaluation, and discovery of optimal sequences within the quantum domain.

\subsection{Demonstration of quantum implementation of MDP}\vspace{-6pt}
After analyzing the classical MDP dynamic, the next step is implementing it into a quantum circuit by encoding the components of the MDP into qubits, allowing for a superposition quantum effect.
\subsubsection{Agent and environment on quantum computer}
\begin{figure*}[ht]
\includegraphics[width=\textwidth]{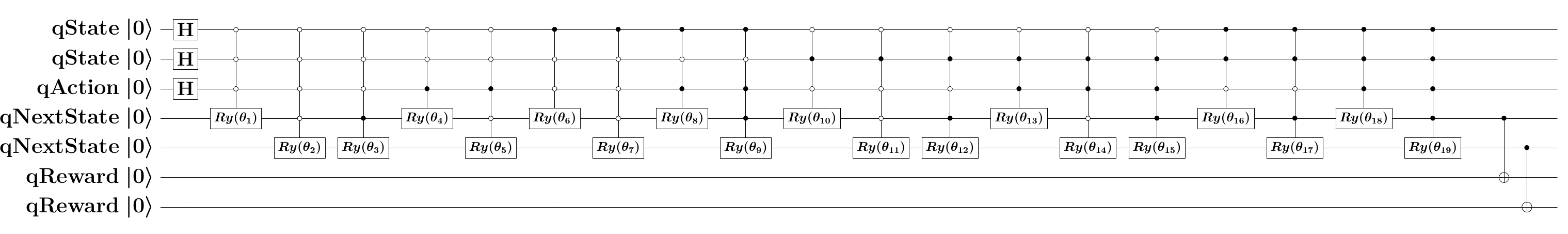}
    \caption{Quantum circuit of the quantum Markov decision process (QMDP) simulating a single interaction between the agent and the environment. The circuit encodes states and actions into qubits, allowing the agent to explore multiple states in superposition. $R_y(\theta)$ gates represent probabilistic state transitions based on the environment’s response to the agent's actions, while CNOT gates implement the reward mechanism, conditioned on the resulting states.}
    \label{fig:one_step}
\end{figure*}

Figure~\ref{fig:one_step} illustrates the quantum Markov decision process (QMDP), where the agent’s actions and the environment’s responses are represented by quantum states. 
The circuit begins by preparing the superpositions of states and actions on quantum registers, representing the starting conditions of the agent-environment interaction. Rotation gates $R_y(\theta)$ represent the state transition probabilities, analogous to how probabilities govern transitions in the classical MDP. These gates parameterize the probability of transitioning between states, but in the quantum domain, they introduce quantum amplitudes instead of classical probabilities, allowing the agent to explore multiple potential states at once, due to quantum superposition. 
The controlled $R_y(\theta)$ rotations apply state transitions conditionally, such that only specific state-action pairs (encoded in the control qubits) activate the rotation on the next state registers.
Mathematically, the rotation angle $\theta$ is determined by the transition probability $P(s'|s,a)$, encoding it into the quantum amplitudes of the next state registers. CNOT gates model the reward mechanism. Specifically, the CNOT gates connect the next state registers to the reward qubits, ensuring that the reward is applied conditionally, depending on the resulting state.

By implementing the classical MDP in the quantum domain, the quantum model allows the agent to explore and evaluate multiple state-action pairs, as well as their potential next states and rewards, simultaneously. By leveraging quantum superposition, this exploration enables the agent to capture the complex dynamics of the environment, including its stochastic transitions and multiple possible outcomes in parallel, which would otherwise require repeated sampling in classical settings. As a result, the quantum model reduces the number of required interactions, leading to improved sample efficiency and potentially enhancing computational performance. Therefore, these transformations enhance the agent’s ability to explore the state space more effectively than classical methods.

\subsubsection{Agent and environment single interaction on quantum computer}
To verify the accuracy of the QMDP circuit, we simulate a single interaction between the agent and the environment on a quantum computer. Using both a state transition heat-map and a quantum sample distribution plot, we demonstrate that the QMDP circuit correctly mirrors the state transitions and rewards of the classical MDP, mentioned in Fig.~\ref{fig:state_transition}.
\begin{figure}[h]
    \includegraphics[width=0.5\textwidth]{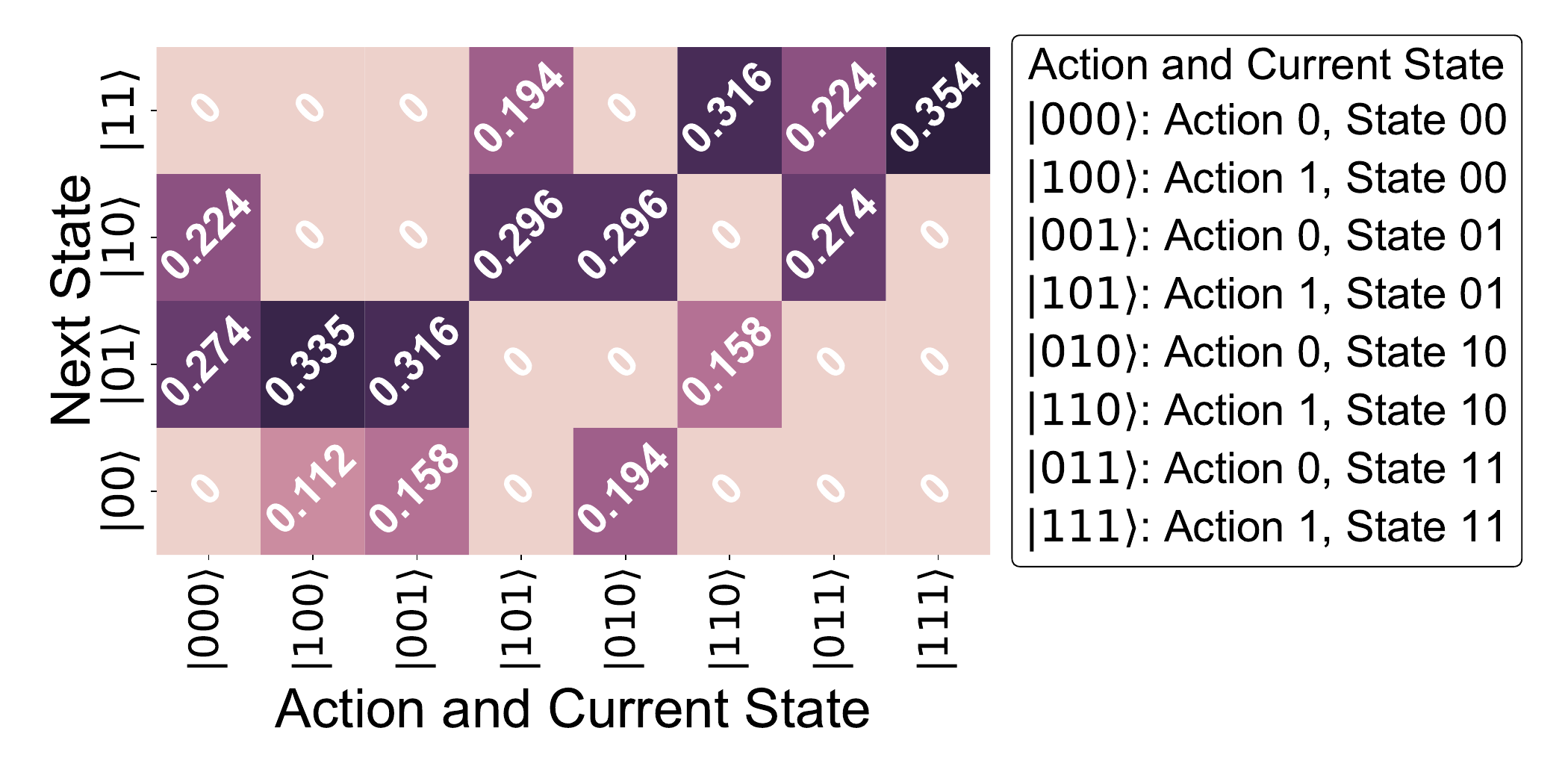}
    \caption{State transition heat-map representing the probabilities of transition from each state-action pair (on the x-axis) to the next state (on the y-axis) within a single agent-environment interaction in QMDP. Darker cells indicate higher transition probabilities.}
    \label{fig:heatmap}
\end{figure}

Figure~\ref{fig:heatmap} presents the state transition heat-map, which captures the probabilities of transition from each current state-action pair (represented along the x-axis) to possible next states (represented along the y-axis) in a single quantum interaction. The agent’s current state and action are encoded as a three-bit binary value on the x-axis, while the resulting next states are on the y-axis. The color intensity of each cell represents the likelihood of each transition, with darker shades indicating higher probabilities. For example, transitioning from the state-action pair $\ket{100}$ (state $s_0$(‘00’) with action $a_1$(‘1’))  has a probability of leading to specific next states, with a peak probability of 0.335 to $s_1$(‘01’). This heat-map confirms that the QMDP circuit preserves the state transition probabilities of the classical MDP.

\begin{figure}[h]
    \includegraphics[width=0.5\textwidth]{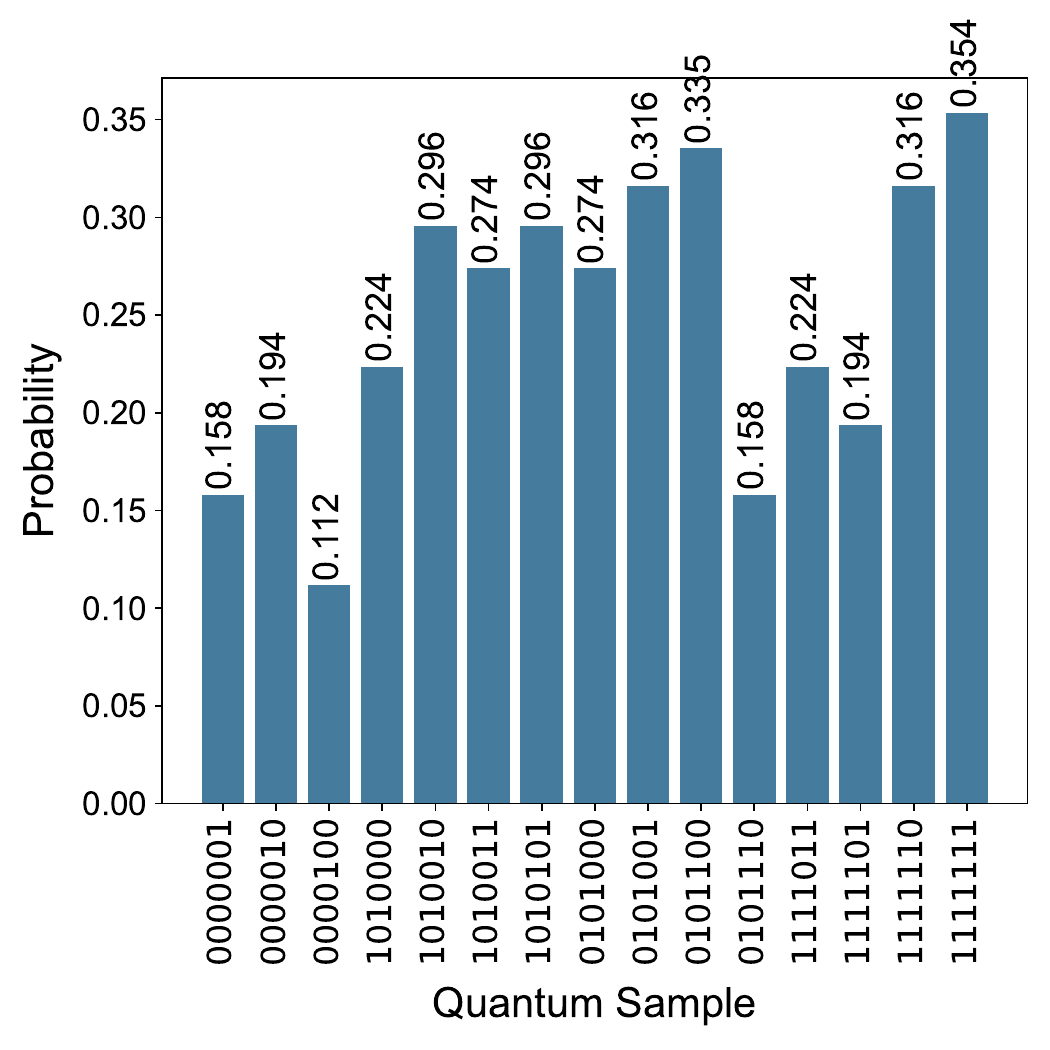}
    \caption{Quantum sample distribution of the QMDP circuit, displaying the probability of measuring each quantum sample after simulating a single agent-environment interaction. The x-axis lists possible quantum samples, while the y-axis indicates their probabilities.}
    \label{fig:sample_distribuition}
\end{figure}
Figure~\ref{fig:sample_distribuition} illustrates the quantum sample distribution of the QMDP circuit, displaying the probability of observing each possible quantum sample after a single execution of the circuit. Each quantum sample in binary notation (x-axis) represents a combination of the agent’s initial state, chosen action, resulting next state and reward sequence encoded in different qubits. Table~\ref{table:qsampledistribution} presents the qubit representations of each quantum sample. Starting from the least significant bits, the first two qubits represent the initial (current) state , followed by the action qubit, then the next two qubits indicate the next state, and finally, the two most significant bits represent the reward. The y-axis indicates the probability of each quantum sample. Peaks in this distribution, quantum samples $\ket{1111111}$ and $\ket{0101100}$ with probabilities 0.354 and 0.335 respectively, highlight high-probability sequences of this interaction. These results validate the QMDP circuit’s accurate representation of both state transition probabilities and reward during a single interaction. 
\begin{table}[h]
\caption{\label{table:qsampledistribution} Qubit representation of quantum samples for a single agent-environment interaction.}
\centering
\renewcommand{\arraystretch}{0.7}   
\begin{ruledtabular}
\begin{tabular}{ccccc}
 Quantum sample & Reward & Next state & Action & Initial state \\
  \hline   \noalign{\vskip 3pt}
     ‘0000001' & 00 & 00 & 0 & 01 \\
     ‘0000010' & 00 & 00 & 0 & 10 \\
     ‘0000100' & 00 & 00 & 1 & 00 \\
     ‘1010000' & 10 & 10 & 0 & 00 \\
     ‘1010010' & 10 & 10 & 0 & 10 \\
     ‘1010011' & 10 & 10 & 0 & 11 \\
     ‘1010101' & 10 & 10 & 1 & 01 \\
     ‘0101000' & 01 & 01 & 0 & 00 \\
     ‘0101001' & 01 & 01 & 0 & 01 \\
     ‘0101100' & 01 & 01 & 1 & 00 \\
     ‘0101110' & 01 & 01 & 1 & 10 \\
     ‘1111011' & 11 & 11 & 0 & 11 \\
     ‘1111101' & 11 & 11 & 1 & 01 \\
     ‘1111110' & 11 & 11 & 1 & 10 \\
     ‘1111111' & 11 & 11 & 1 & 11 \\
\end{tabular}
\end{ruledtabular}
\end{table}
Together, these figures validate that the QMDP circuit in Fig.~\ref{fig:one_step} accurately reproduces the transition probabilities of the classical MDP, in Fig.~\ref{fig:state_transition}, within a quantum framework. The state transition heat-map (Fig.~\ref{fig:heatmap}) shows that the quantum circuit’s state transition aligns with classical state transition probabilities, while the quantum sample distribution (Fig.~\ref{fig:sample_distribuition}) confirms that high-probability sequences are prominently represented in the quantum outcomes. This consistency demonstrates that the QMDP circuit can accurately simulate the agent-environment interaction, capturing the dynamics of the classical MDP while leveraging quantum superposition.
\vspace{-5pt}

\subsection{Demonstration of agent-environment quantum interactions over multiple time steps}
\begin{figure*}[ht]
    \includegraphics[width=\textwidth]{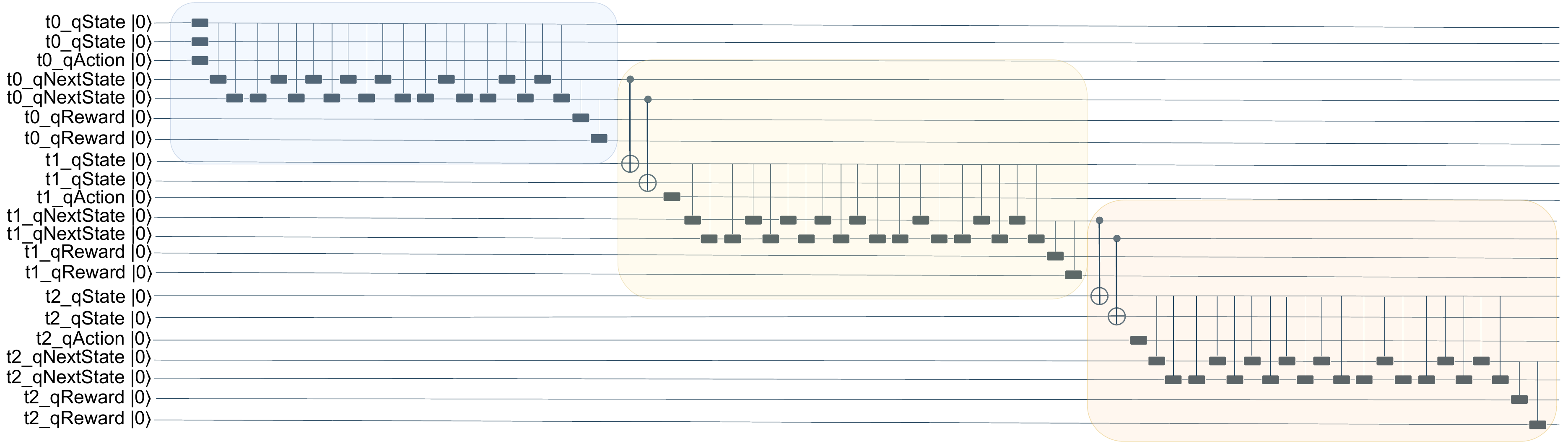}
    \caption{Quantum circuit implementation of agent-environment interactions across 3 time steps (t = 0, 1, 2). Each colored block represents a single time step, with qubits for states (qState), actions (qAction), next states (qNextState), and rewards (qReward). CNOT gates represent state transitions between the interactions.}
    \label{fig:two_state_block}
\end{figure*}

\vspace{-6pt}
In this section, we extend the QMDP model to simulate multiple agent-environment interactions over 3 consecutive time steps, focusing on how the state evolves based on repeated interactions of the agent and the environment. Figure~\ref{fig:two_state_block} illustrates the QMDP circuit for this 3-time step process, where each time step (labeled $t0, t1$, and $t2$) represents a unique interaction between the agent and the environment. To model the state transitions between time steps, CNOT gates are employed at each step. These gates update the state qubits of the following time step conditionally, depending on the evolution of the new state in response to the prior interaction of the agent with the environment. This configuration allows the QMDP circuit to evaluate not just individual state transitions, but also the cumulative effect of the agent’s interactions over 3 time steps, effectively capturing the rewards across multiple agent and environment interactions. 

\begin{figure*}[ht]
    \includegraphics[width=\textwidth]{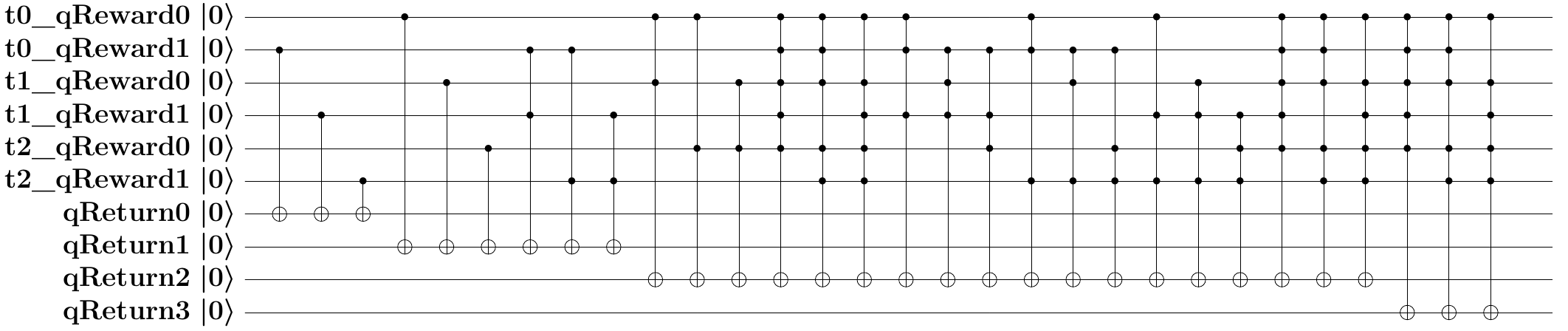}
    \caption{Quantum circuit for return calculation in the QMDP. The process simulates the overall outcome of the agent-environment interactions by summing rewards from all time steps into the return value by means of a quantum arithmetic.}
    \label{fig:return_circuit} 
\end{figure*}

After the agent-environment interaction across 3 time steps, the next step is to calculate the return, which represents the cumulative reward the agent accumulates over these time steps. In classical RL, the return is typically the discounted sum of rewards from each time step, weighted by a discount factor. However, in this quantum implementation, for simplicity, we set the discount factor to 1, meaning each reward are equally treated, allowing for a direct accumulation of rewards across these time steps. This simplification allows us to focus on the mechanics of quantum return calculation, as demonstrated in Fig.~\ref{fig:return_circuit}.
In the return circuit of Fig.~\ref{fig:return_circuit}, qubits corresponding rewards from each time step (denoted as t0${\_}$qReward0, t0${\_}$qReward1, t1${\_}$qReward0, t1${\_}$qReward1, and t2${\_}$qReward0, t2${\_}$qReward1) are processed into quantum registers which are labeled as qReturn0, qReturn1, qReturn2 and qReturn3, that store the cumulative return. At each time step, the rewards are conditionally added to the return qubits, ensuring a straightforward accumulation of rewards across time steps. The CNOT, Toffoli, and multiple-controlled x-gates link the reward qubits from each time step to the return registers, which sum the rewards obtained across these interactions. The use of quantum arithmetic to compute the return enables the quantum model to reflect the cumulative benefits experienced by the agent, just as in the classical RL.

\begin{figure*}[ht]
    \includegraphics[width=\textwidth]{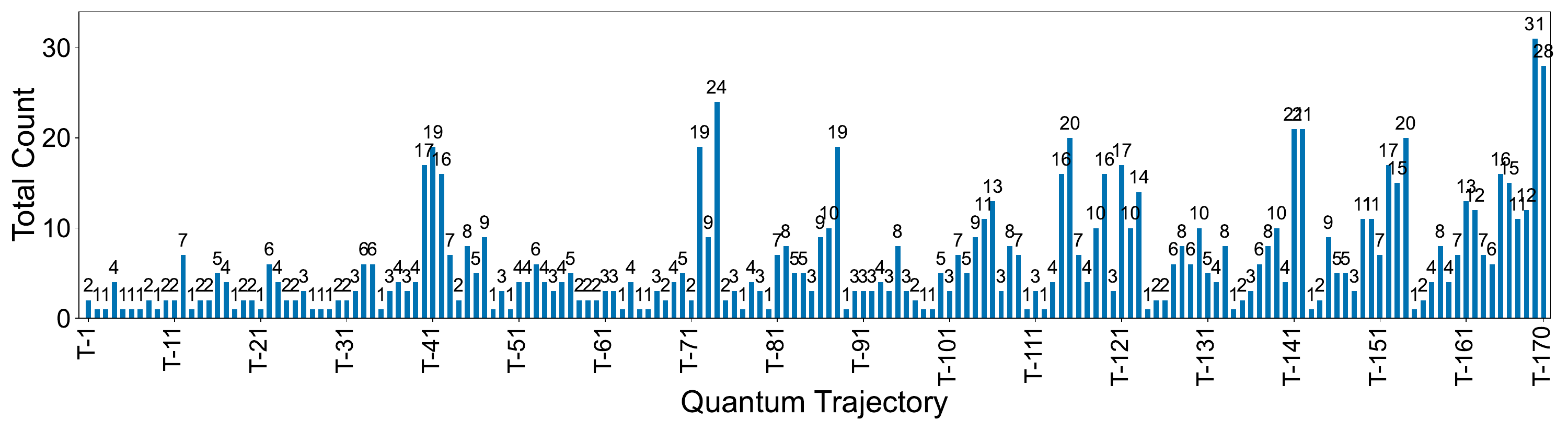}
    \caption{Distribution of quantum trajectories in the QMDP for 3 time steps. The x-axis shows trajectory numbers (see Table~\ref{table:classical_trajectories} in Appendix \ref{appendixA} for the corresponding quantum trajectories.) and the y-axis shows the total count of occurrences for each trajectory. Higher bars indicate more frequent trajectories, providing insight into the agent-environment interaction outcomes over multiple time steps.  }
    \label{fig:classical}
\end{figure*}

Now, we present the classical measurement outcomes obtained from simulating the agent-environment interactions over 3 time steps. The classical measurement phase provides insight into the specific quantum trajectories generated by the QMDP circuit, reflecting the different paths the agent took based on its actions and the corresponding state transitions. In this demonstration, we used IBM Qiskit's Aer simulator to perform this measurement (see Appendix \ref{appendixC} for detailed specifications). Figure~\ref{fig:classical} shows the quantum trajectory distribution over 3 time steps. Table~\ref{table:classical_trajectories} (see Appendix \ref{appendixA}) presents the list of quantum trajectories along with their corresponding trajectory numbers. Each quantum trajectory is represented as a binary string, which corresponds to the sequences of state-action-next state-reward and the return which are observed during the simulation of the interactions over 3 time steps. The y-axis displays the total count of occurrences for each trajectory. Trajectories that occur more frequently are represented by higher bars, showing which paths the agent most followed during these interactions.

The distribution reveals that some trajectories appear significantly more often than others, which demonstrates that the agent-environment interaction can result in different outcomes based on the stochastic nature of the environment and the agent’s chosen action. The diversity of these trajectories highlights the range of possible decision paths the agent can take over multiple time steps in the quantum model. This distribution allows us to assess the behavior of the QMDP system and identify the most probable trajectories that emerged during this simulation.
\vspace{-5pt}

\subsection{Demonstration of quantum trajectories search}\vspace{-6pt}
In this section, we present the results of implementing Grover's search algorithm to identify optimal quantum trajectories within the quantum reinforcement learning (QRL) framework outlined in the method section. The objective of this demonstration is to efficiently search through the set of possible quantum trajectories generated from agent-environment interactions to find those that maximize the overall return, thereby guiding the discovery of optimal policies for the QRL agent. Each trajectory is composed of a sequence of states, actions, next states, rewards, and a return value, and Grover’s search algorithm is employed to efficiently identify those trajectories with the highest return from a large set of possible agent-environment interactions.

To evaluate the effectiveness of Grover's search in identifying high-return trajectories, we analyze two specific scenarios. In the first scenario, the agent begins from a fixed initial state $s_0$ and seeks to terminate at state $s_3$, while maximizing the total return over 3 time steps.  This case represents a scenario where the agent's initial conditions are predetermined, allowing us to evaluate the performance of Grover's search algorithm to find high-return trajectories under fixed initial conditions. In the second scenario, we consider the initial condition by allowing the agent to start from any state with the same probability within the state space and again terminate at state $s_3$, aiming to maximize the return.  This scenario offers a broader search space, enabling us to assess the flexibility and efficiency of Grover's search in handling less restrictive conditions where the agent’s starting state can vary.

The following provides a detailed analysis of the results for first scenario, illustrating the performance of Grover’s search algorithm in identifying the optimal quantum trajectories that yield the maximum return. These results reveal the potential of quantum algorithms to enhance trajectory search processes in reinforcement learning tasks. This demonstration not only illustrates the application of Grover’s search algorithm but also highlights the significant advantages gained from implementing a classical Markov decision process (MDP) into the quantum domain. By manipulating the quantum version of MDP, we can efficiently search through possible trajectories and state-action pairs using Grover’s search in only one call to the oracle. This quantum-enhanced framework provides a substantial computational advantage over classical methods, which would typically require more computational resources to explore the same search space.
\begin{figure*}[ht]
    \includegraphics[width=\textwidth]{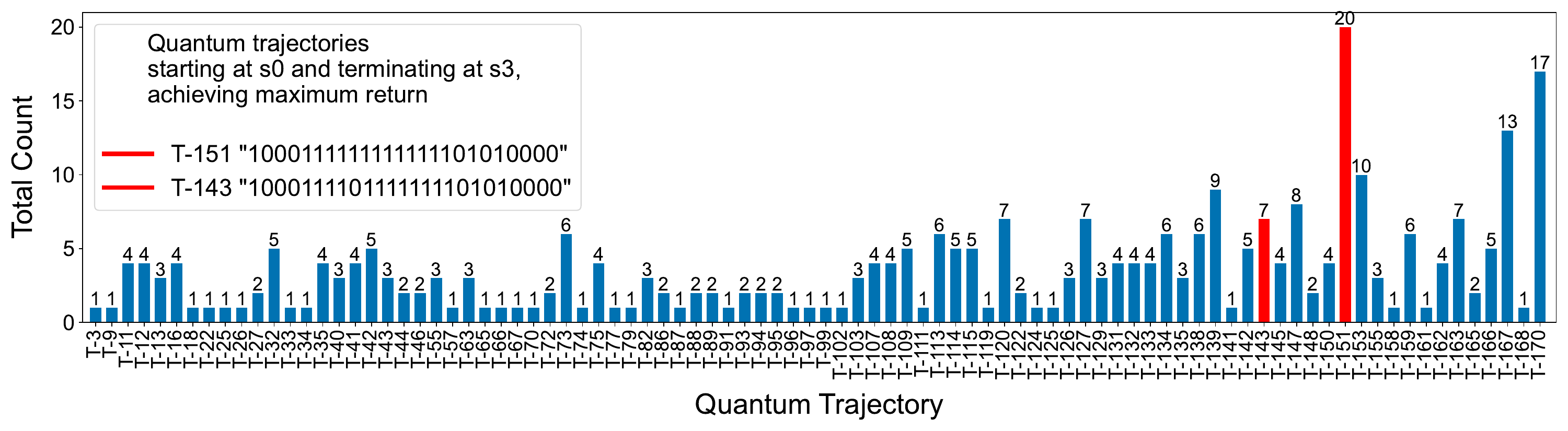}
    \caption{Distribution of quantum trajectories after executing Grover's algorithm to search the trajectories starting at $s_0$ and terminating at $s_3$ in a QMDP over 3 time steps. Each trajectory number represents unique quantum sequences. The highest-return trajectories for this scenario are highlighted in red. The most frequent trajectory was sampled 20 times, indicating its optimality in this scenario.}
    \label{fig:starting_s0}
\end{figure*}

In the first case, we examine the scenario where the agent starts from a fixed initial state $s_0$ and terminates at state $s_3$, aiming to achieve the maximum return over three time steps. The result of this demonstration is visualized in Fig.~\ref{fig:starting_s0}, which shows the distribution of quantum trajectories identified by Grover’s search algorithm, as well as their respective counts based on how frequently each trajectory was sampled. The numbers along the x-axis represent distinct quantum trajectories numbers, each representing a unique sequences of states, actions, next states, rewards and return over 3 time steps (see Table~\ref{table:classical_trajectories} in Appendix \ref{appendixA} for the corresponding quantum trajectories). The bar illustrates the total count of each trajectory sampled during the measurement process after executing Grover’s search. 

Based on the classical measurement of trajectory distribution for the QMDP over 3 time steps which is illustrated in Fig.~\ref{fig:classical}, there are a total of 11 possible trajectories where the agent starts at $s_0$ and successfully terminates at $s_3$, as detailed in Table~\ref{table:trajectories}. Among these, two trajectories achieve the maximum return (‘1000’).
\begin{table*}[ht]
\caption{\label{table:trajectories}Possible trajectories where the agent starts at $s_0$ and terminates at $s_3$.}
\centering
\renewcommand{\arraystretch}{0.7}  
\begin{ruledtabular}
\begin{tabular}{ccccc}
 Quantum trajectory & Return & Time step t=2  & Time step t=1 & Time step t=0 \\
  \hline \noalign{\vskip 3pt}
 ‘0101111110101010010101000' & 0101 & 1111101 & 0101001 & 0101000 \\
 ‘0101111110101010010101100' & 0101 & 1111101 & 0101001 & 0101100 \\
 ‘0110111111010101010101000' & 0110 & 1111110 & 1010101 & 0101000 \\
 ‘0110111111010101010101100' & 0110 & 1111110 & 1010101 & 0101100 \\
 ‘0111111101111111010101000' & 0111 & 1111011 & 1111101 & 0101000 \\
 ‘0111111101111111010101100' & 0111 & 1111011 & 1111101 & 0101100 \\
 ‘0111111111010100101010000' & 0111 & 1111110 & 1010010 & 1010000 \\
 ‘0111111111111111010101100' & 0111 & 1111111 & 1111101 & 0101100 \\
 ‘0111111111111111010101000' & 0111 & 1111111 & 1111101 & 0101000 \\
 ‘1000111111111111101010000' & 1000 & 1111111 & 1111110 & 1010000 \\
 ‘1000111101111111101010000' & 1000 & 1111011 & 1111110 & 1010000 
\end{tabular}
\end{ruledtabular}
\end{table*}
In Table~\ref{table:trajectories}, each quantum sample for a given time step is represented by a structured sequence of qubits. 
The qubit representation for each time step is detailed in Table~\ref{table:qsampledistribution}.

As shown in Fig.~\ref{fig:starting_s0}, Grover’s algorithm successfully identifies these optimal trajectories, marked in red, which correspond to paths that achieve the highest return. The two optimal trajectories, ‘‘1000111111111111101010000’’ and ‘‘1000111101111111101010000’’, represent the most favorable paths under this condition, where the agent successfully navigates from $s_0$ to $s_3$ with the highest return. The trajectory ‘‘1000111111111111101010000’’ was sampled 20 times, indicating that it is the most frequently identified optimal trajectory in the quantum search process. Starting from the least significant bit, this trajectory unfolds as follows: at the initial time step $t=0$, the agent begins at $s_0$(‘00’), takes action $a_0$ (‘0’), and reaches $s_2$ (‘10’) with a reward $r_2$ (‘10’), representing ‘‘1010000’’. At time step $t=1$, while at state $s_2$ (‘10’), the agent takes action $a_1$ (‘1’), moving to the next state $s_3$ (‘11’) with reward $r_3$ (‘11’), described as ‘‘1111110’’. Finally, at time step $t=2$, the agent takes action $a_1$ (‘1’) again, remaining at state $s_3$ (‘11’) and receiving the reward $r_3$ (‘11’) which is represented as ‘‘111111’’. This sequence generates the maximum return of $8$ (‘1000’) for the entire trajectory. The second optimal trajectory, ‘‘1000111101111111101010000’’, was sampled 7 times, demonstrating another high-return path for this specified condition. 
According to these results, we can conclude that the optimal action for each state under this scenario is as follows: for state $s_0$ (‘00’), the optimal action is $a_0$ (‘0’), and for states $s_2$ (‘10’) and $s_3$ (‘11’), the optimal action is $a_1$ (‘1’). Grover's search finds these optimal solutions with only one query to the oracle. This highlights the computational power of quantum algorithms in reinforcement learning tasks.

To compare these quantum results with a classical RL approach, we conducted an analysis using Q-learning on an equivalent MDP dynamic. For this comparison, we set a discount factor of 1 to allow us to evaluate the consistency and effectiveness of the trajectories identified by Grover’s search with those found through classical Q-learning. In an initial step of the Q-learning process, the Q-table was set to zero and updated through iterative runs of the Q-learning algorithm until it converged toward the optimal policy. Table~\ref{table:qvaluestartings0} presents the resulting Q-table, which contains the Q-values learned for an environment with four states ($s_0, s_1, s_2, s_3$) and two actions ($a_0, a_1$). Each cell in the table contains the Q-value for a state-action pair, which represents the expected cumulative reward when taking a respective action in that given state. For each state, the action with the highest Q-value is highlighted in yellow, indicating the optimal action to maximize expected rewards. In state $s_0$, action $a_0$ has the highest Q-value (38.02), making it optimal. In states $s_1$, $s_2$, and $s_3$, action $a_1$ is optimal, with Q-values of 47.89, 54.28, and 59.54, respectively. 
\begin{table}[h!]
\caption{\label{table:qvaluestartings0}Learned Q-values for each state-action pair in a 4-state environment. Highlighted values indicate the optimal action for each state based on Q-learning results.}
\begin{ruledtabular}
\centering
\renewcommand{\arraystretch}{0.7}  
\begin{tabular}{ccc}
 & $a_0$ & $a_1$  \\
  \hline
  \noalign{\vskip 3pt}
    $s_0$ & \cellcolor{yellow} 38.02 & 30.22 \\ 
    $s_1$ & 37.70 & \cellcolor{yellow} 47.89 \\ 
    $s_2$ & 42.20 & \cellcolor{yellow} 54.28 \\ 
    $s_3$ & 56.58 & \cellcolor{yellow} 59.54 \\ 
\end{tabular}
\end{ruledtabular}
\end{table}

\begin{figure}[t]
   \vspace{10pt}
    \includegraphics[scale = 1.5, width=0.5\textwidth]{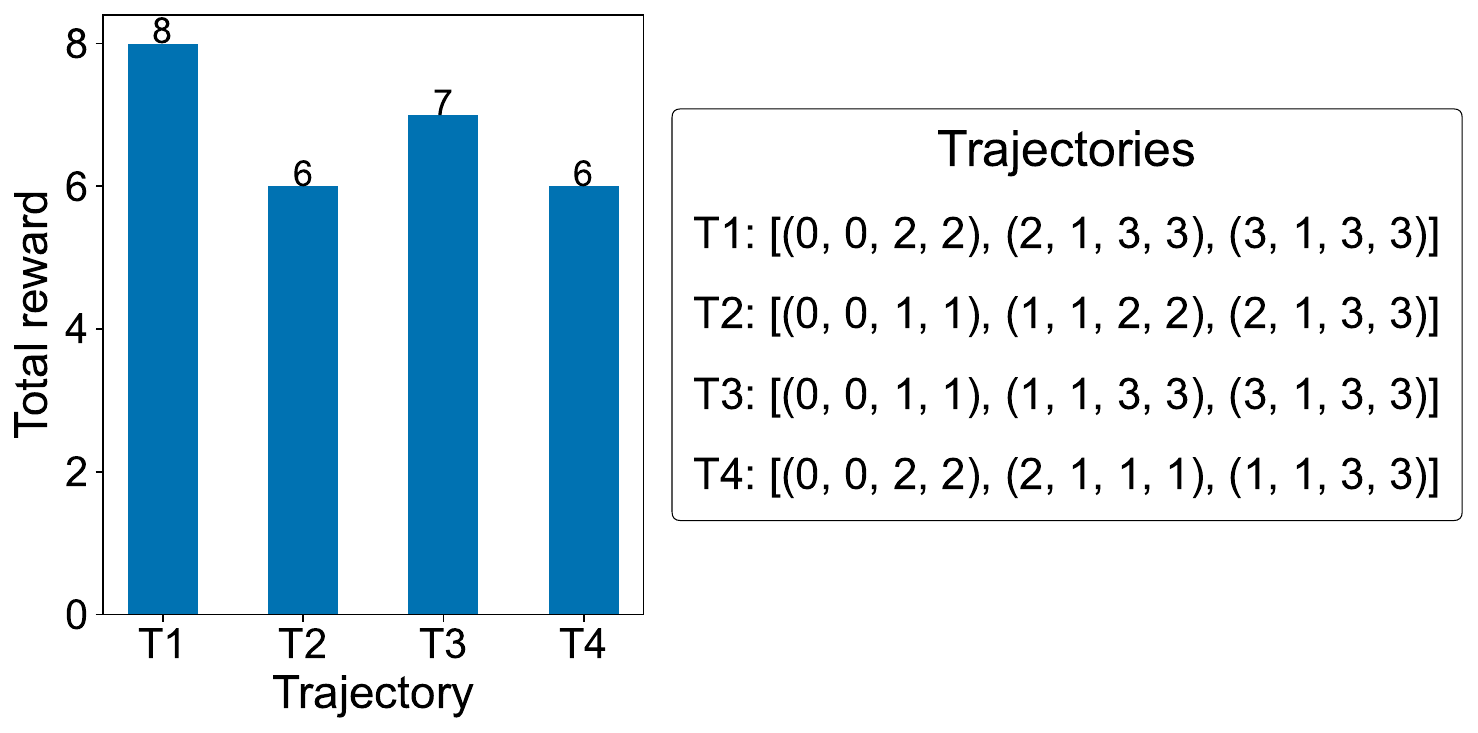}
    \caption{Comparison of total rewards for 4 unique trajectories from $s_0$ to $s_3$ over 3 time steps in a classical Q-learning. Trajectory $T1$ achieves the highest reward of 8, followed by $T3$ with a reward of 7, and both $T2$ and $T4$ with rewards of 6. The legend details the state-action sequences for each trajectory.}
    \label{fig:reward_s0}
\end{figure}

\begin{figure}[t]
    \includegraphics[scale = 1.5, width=0.5\textwidth]{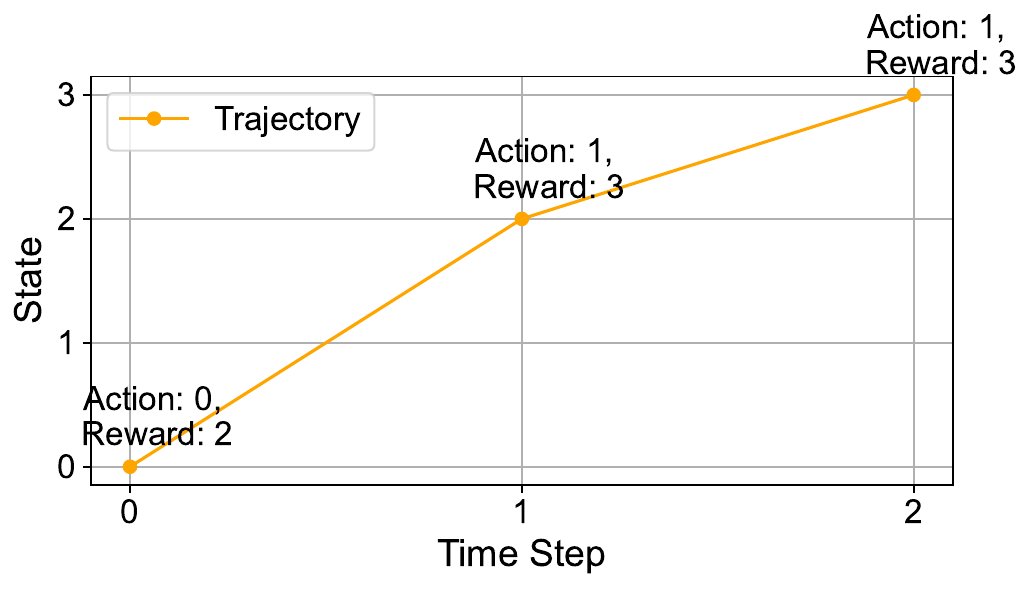}
    \caption{Optimal trajectory plot for an agent transitioning from $s_0$ to $s_3$ over 3 time steps in a classical Q-learning. Each action and reward are labeled along the path, showing a total cumulative reward of 8 for this trajectory.}
    \label{fig:optimal_s0}
\end{figure}

After generating the Q-table, we used the learned Q-values to evaluate potential trajectories by running the Q-learning algorithm over 100 trials, each consisting of 3 time steps. This process resulted in four unique trajectories, each with an associated total reward, as shown in Fig.~\ref{fig:reward_s0}. Figure~\ref{fig:reward_s0} provides a comparison of total rewards for these four trajectories, beginning at state $s_0$ and terminating at state $s_3$. The y-axis represents the total reward, while the x-axis lists the different trajectories. Each bar represents a distinct trajectory, labeled $T1, T2, T3,$ and $T4$, with the total reward achieved by each trajectory shown above the corresponding bar. Among these, trajectory $T1$ achieves the highest total reward of $8$, identifying it as the optimal path. In $T1$, the agent starts at $s_0$, takes action $a_0$ to reach $s_2$ and receive $r_2$ which is represented as ‘‘(0,0,2,2)’’, then takes action $a_1$ to move to $s_3$ with $r_3$ described as ‘‘(2,1,3,3)’’, and finally takes $a_1$ again to remain in $s_3$ representing ‘‘(3,1,3,3)’’. This sequence results in the highest total reward among all trajectories, indicating that it is the most rewarding path available within the constraints.
Figure~\ref{fig:optimal_s0} provides a detailed visualization of the optimal trajectory over 3 time steps, showing the sequence of states, actions, and corresponding rewards. The plot shows the agent's movement through states across discrete time steps, along with the actions taken and rewards obtained at each step. 

Upon comparison, the optimal trajectory obtained from the classical Q-learning process ($T1$) was found to match the result produced by Grover’s search algorithm applied to the quantum MDP (‘‘1000111111111111101010000’’). This consistency between the quantum and classical methods demonstrates that Grover’s quantum search effectively replicates the optimal solution achievable by traditional Q-learning in this context. Such alignment reinforces the potential of quantum search algorithms in solving MDPs by efficiently identifying optimal policies comparable to those found with classical RL methods. The trajectory search for the second scenario using Grover's algorithm is presented in Appendix \ref{appendixB}.
\vspace{-5pt}

\subsection{Advantages of quantum RL over classical approaches}\vspace{-6pt}
A core benefit of our work lies in its ability to execute the entire reinforcement learning process within the quantum domain.
The integration of quantum information formalism into reinforcement learning presents several significant advantages over classical approaches.

By representing both the agent and environment as quantum components and leveraging quantum computations, our quantum framework enables the quantum agent to evaluate numerous interaction sequences in a single execution, significantly reducing the number of interactions required to explore the environment. As a result, the quantum approach achieves much greater sample efficiency compared to classical methods, which typically process one interaction sequence at a time and require many iterations to converge on an optimal policy.

Moreover, the implementation of Grover’s search algorithm enables efficient trajectory search with a single oracle call. This quantum-enhanced approach significantly accelerates the identification of high-return trajectories while avoiding the iterative evaluations required by classical methods. In our demonstration, the trajectories identified by the quantum approach match those found by classical Q-learning but are achieved with substantially lower computational cost. This highlights the effectiveness of Grover’s algorithm in delivering significant speedups in trajectory search compared to classical reinforcement learning algorithms. By integrating quantum superposition and Grover’s search, the proposed framework achieves both improved sample efficiency and substantial computational advantages over classical reinforcement learning.
\vspace{-5pt}

\section{Discussion}\label{sec:discussion}\vspace{-6pt}
The results validate that the quantum implementation of the Markov decision process (QMDP) not only aligns with classical MDP dynamics but also effectively leverages quantum properties to enhance performance. Our quantum model maintains the core structure of state transitions and action choices as observed in the classical MDP and ensures that the decision-making of QMDP remain consistent with that of classical MDP. This validates that our quantum approach can successfully replicate the behavior of classical MDP while introducing potential improvements in speed and computational efficiency. Another key observation is that with only 3 time steps, the quantum agent can effectively explore the possible trajectories and observe the optimal policies for each state. The quantum agent's ability for simultaneous exploration of multiple trajectories, resulting in a faster and more efficient determination of the optimal policy compared to traditional algorithms. Furthermore, the use of Grover’s search algorithm within our quantum MDP framework demonstrates that a single call to the oracle is sufficient to identify the desired trajectories. Grover’s algorithm offers a speedup, allowing the quantum agent to perform optimal trajectory searches with substantially fewer computational steps. This efficient solution highlights an improvement in speed and computational resources compared to classical search algorithms. These advantages highlight the potential of quantum computing in complex search and optimization processes, providing faster solutions with reduced computational resources.
\vspace{-5pt}

\section{Conclusion}\label{sec:conclusion}\vspace{-6pt}
In this work, we have presented a quantum implementation of a classical Markov decision process (MDP) entirely within the quantum framework, modeling both the agent and environment as quantum components, as well as their interactions are exclusively through quantum processes. Furthermore, the trajectory search is also conducted entirely within a quantum framework to reduce the computational complexity of trajectory optimization.  By demonstrating a purely quantum approach to the agent-environment interactions and trajectory optimization, we achieve a complete quantum realization of reinforcement learning (QRL) without any reliance on classical computations, offering an efficient alternative to classical reinforcement learning tasks.

This work serves as a foundational step in advancing the field of quantum reinforcement learning (QRL). It provides a concrete basis for future research aimed at scaling quantum-native RL systems, exploring more complex environments, and integrating with evolving quantum hardware platforms. Beyond its theoretical contributions, the proposed framework demonstrates strong potential for real-world applications.
In autonomous driving, quantum MDP enables simultaneous evaluation of multiple driving trajectories, while Grover’s search accelerates optimal policy discovery, leading to faster decision-making for collision avoidance and route planning. In personalized healthcare, patient conditions and treatments can be modeled as MDPs, and the QRL framework evaluates multiple treatment plans in parallel, with Grover’s search identifying those with the highest reward for faster selection of effective therapies. In financial portfolio management, our approach enables parallel exploration of investment strategies, with Grover’s search rapidly identifying high-return paths, offering a significant computational advantage for real-time trading environments.
These real-world scenarios illustrate how the proposed framework provides practical applicability and computational advantages, establishing a foundation for future advances in quantum reinforcement learning applications.

Future research could be extended by exploring more complex MDPs with larger state and action spaces, addressing scalability challenges by optimizing qubit usage. 
In our current implementation of Grover’s search, we assume prior knowledge of the maximum return and construct the quantum oracle accordingly. A promising direction for future work is to develop quantum approaches capable of searching for unknown optimal returns without this prior knowledge. Additionally, since qubit usage increases significantly when running over multiple time steps, investigating techniques for reusing qubits could help mitigate resource demands. 
Exploring alternative quantum search algorithms beyond Grover’s could further enhance efficiency in various decision-making scenarios. 
These advancements could expand the applicability of quantum MDPs and drive progress in quantum reinforcement learning for complex, resource-intensive tasks.
\vspace{-5pt}

\begin{acknowledgments}\vspace{-6pt}
This work was supported by the Center of Innovations for Sustainable Quantum AI (JST Grant Number JPMJPF2221), JSPS KAKENHI Grant Number JP24K20843, and the Ministry of Education, Culture, Sports, Science and Technology (MEXT) of Japan. We acknowledge the use of IBM Quantum services for this work. The views expressed are those of the authors, and do not reflect the official policy or position of IBM or the IBM Quantum team.
\end{acknowledgments}
\newpage

\appendix
\onecolumngrid 
\section{Quantum Trajectories for Agent-Environment Interactions Over 3 Time Steps}
\label{appendixA}
\vspace{-20pt}
\begin{table*}[ht]
\caption{List of quantum trajectories with their corresponding trajectory numbers. Each quantum trajectory represents a unique sequence of interactions and the return over 3 time steps.}
\label{table:classical_trajectories}
\centering
\begin{minipage}{0.33\linewidth}
    \centering
    \renewcommand{\arraystretch}{0.55}
    \begin{ruledtabular}
    \begin{tabular}{@{}c@{\hskip -0.1cm}c@{}}
        Trajectory no.& Quantum trajectory\\
        \hline
        \noalign{\vskip 5pt} 
             T-1 & ‘0001000000101010000000001' \\
             T-2 & ‘0001000000101010000000100' \\
             T-3 & ‘0001000000101011000000001' \\
             T-4 & ‘0001000000101011000000010' \\
             T-5 & ‘0001000010000000010101100' \\
             T-6 & ‘0001010100000001000000001' \\
             T-7 & ‘0001010100000001000000010' \\
             T-8 & ‘0001010110000001000000001' \\
             T-9 & ‘0001010110000001000000010' \\
             T-10 & ‘0010000000101010010101000' \\
             T-11 & ‘0010000000101010010101001' \\
             T-12 & ‘0010000000101010010101100' \\
             T-13 & ‘0010000000101010010101110' \\
             T-14 & ‘0010000001010100000000010' \\
             T-15 & ‘0010000010000000101010010' \\
             T-16 & ‘0010010100000000010101001' \\
             T-17 & ‘0010010100000000010101100' \\
             T-18 & ‘0010010100000000010101110' \\
             T-19 & ‘0010010100101010000000001' \\
             T-20 & ‘0010010100101010000000010' \\
             T-21 & ‘0010010100101010000000100' \\
             T-22 & ‘0010010100101011000000001' \\
             T-23 & ‘0010010100101011000000010' \\
             T-24 & ‘0010010100101011000000100' \\
             T-25 & ‘0010010110000000010101000' \\
             T-26 & ‘0010010110000000010101001' \\
             T-27 & ‘0010010110000000010101100' \\
             T-28 & ‘0010010110000000010101110' \\
             T-29 & ‘0010101000000001000000010' \\
             T-30 & ‘0011000000101011101010000' \\
             T-31 & ‘0011000000101011101010010' \\
             T-32 & ‘0011000001010101010101000' \\
             T-33 & ‘0011000001010101010101001' \\
             T-34 & ‘0011000001010101010101100' \\
             T-35 & ‘0011000001010101010101110' \\
             T-36 & ‘0011010100000000101010000' \\
             T-37 & ‘0011010100000000101010010' \\
             T-38 & ‘0011010100000000101010011' \\
             T-39 & ‘0011010100000000101010101' \\
             T-40 & ‘0011010100101010010101000' \\
             T-41 & ‘0011010100101010010101001' \\
             T-42 & ‘0011010100101010010101100' \\
             T-43 & ‘0011010100101010010101110' \\
             T-44 & ‘0011010110000000101010000' \\
             T-45 & ‘0011010110000000101010010' \\
             T-46 & ‘0011010110000000101010011' \\
             T-47 & ‘0011010110000000101010101' \\
             T-48 & ‘0011010111010100000000100' \\
             T-49 & ‘0011101000000000010101001' \\
             T-50 & ‘0011101000000000010101110' \\
             T-51 & ‘0011101010101010000000001' \\
             T-52 & ‘0011101010101010000000010' \\
             T-53 & ‘0011101010101011000000001' \\
             T-54 & ‘0011101010101011000000010' \\
             T-55 & ‘0100000001010100101010000' \\
             T-56 & ‘0100000001010100101010010' \\
             T-57 & ‘0100000001010100101010011' \\
    \end{tabular}
    \end{ruledtabular}
\end{minipage}%
\hspace{0.005\linewidth}
\begin{minipage}{0.33\linewidth}
    \centering
    \renewcommand{\arraystretch}{0.55}
    \begin{ruledtabular}
    \begin{tabular}{@{}c@{\hskip -0.1cm}c@{}}
        Trajectory no.&Quantum trajectory\\
        \hline
        \noalign{\vskip 5pt} 
         T-58 & ‘0100000001010100101010101' \\
         T-59 & ‘0100010100101011101010000' \\
         T-60 & ‘0100010100101011101010010' \\
         T-61 & ‘0100010100101011101010011' \\
         T-62 & ‘0100010100101011101010101' \\
         T-63 & ‘0100010111010101010101000' \\
         T-64 & ‘0100010111010101010101001' \\
         T-65 & ‘0100010111010101010101100' \\
         T-66 & ‘0100010111010101010101110' \\
         T-67 & ‘0100101000000000101010011' \\
         T-68 & ‘0100101000000000101010101' \\
         T-69 & ‘0100101001010100000000001' \\
         T-70 & ‘0100101001010100000000010' \\
         T-71 & ‘0100101001010100000000100' \\
         T-72 & ‘0100101010101010010101000' \\
         T-73 & ‘0100101010101010010101001' \\
         T-74 & ‘0100101010101010010101100' \\
         T-75 & ‘0100101010101010010101110' \\
         T-76 & ‘0100111110101010000000010' \\
         T-77 & ‘0100111110101011000000001' \\
         T-78 & ‘0100111110101011000000010' \\
         T-79 & ‘0101000001010100111111011' \\
         T-80 & ‘0101000001010100111111101' \\
         T-81 & ‘0101000001010100111111110' \\
         T-82 & ‘0101000001010100111111111' \\
         T-83 & ‘0101010111010100101010010' \\
         T-84 & ‘0101010111010100101010011' \\
         T-85 & ‘0101010111010100101010101' \\
         T-86 & ‘0101101001010101010101000' \\
         T-87 & ‘0101101001010101010101001' \\
         T-88 & ‘0101101001010101010101100' \\
         T-89 & ‘0101101001010101010101110' \\
         T-90 & ‘0101101010101011101010010' \\
         T-91 & ‘0101101010101011101010011' \\
         T-92 & ‘0101101010101011101010101' \\
         T-93 & ‘0101111110101010010101000' \\
         T-94 & ‘0101111110101010010101001' \\
         T-95 & ‘0101111110101010010101100' \\
         T-96 & ‘0101111111010100000000001' \\
         T-97 & ‘0101111111010100000000010' \\
         T-98 & ‘0101111111010100000000100' \\
         T-99 & ‘0110010111010100111111011' \\
         T-100 & ‘0110010111010100111111101' \\
         T-101 & ‘0110010111010100111111110' \\
         T-102 & ‘0110010111010100111111111' \\
         T-103 & ‘0110101001010100101010000' \\
         T-104 & ‘0110101001010100101010010' \\
         T-105 & ‘0110101001010100101010011' \\
         T-106 & ‘0110101001010100101010101' \\
         T-107 & ‘0110101001111111010101000' \\
         T-108 & ‘0110101001111111010101001' \\
         T-109 & ‘0110101001111111010101100' \\
         T-110 & ‘0110111110101011101010000' \\
         T-111 & ‘0110111110101011101010011' \\
         T-112 & ‘0110111110101011101010101' \\
         T-113 & ‘0110111111010101010101000' \\
         T-114 & ‘0110111111010101010101001' \\
    \end{tabular}
    \end{ruledtabular}
\end{minipage}%
\hspace{0.005\linewidth}
\begin{minipage}{0.325\linewidth}
    \vspace{0pt} 
    \centering
    \renewcommand{\arraystretch}{0.55}
    \begin{ruledtabular}
    \begin{tabular}{@{}c@{\hskip -0.1cm}c@{}}
        Trajectory no.&Quantum trajectory\\
        \hline
        \noalign{\vskip 5pt} 
         T-115 & ‘0110111111010101010101100' \\
         T-116 & ‘0111101001010100111111011' \\
         T-117 & ‘0111101001010100111111101' \\
         T-118 & ‘0111101001010100111111110' \\
         T-119 & ‘0111101001010100111111111' \\
         T-120 & ‘0111101001111111101010000' \\
         T-121 & ‘0111101001111111101010010' \\
         T-122 & ‘0111101001111111101010011' \\
         T-123 & ‘0111101001111111101010101' \\
         T-124 & ‘0111111101111111010101000' \\
         T-125 & ‘0111111101111111010101001' \\
         T-126 & ‘0111111101111111010101100' \\
         T-127 & ‘0111111111010100101010000' \\
         T-128 & ‘0111111111010100101010010' \\
         T-129 & ‘0111111111010100101010011' \\
         T-130 & ‘0111111111010100101010101' \\
         T-131 & ‘0111111111111111010101000' \\
         T-132 & ‘0111111111111111010101001' \\
         T-133 & ‘0111111111111111010101100' \\
         T-134 & ‘0111111111111111010101110' \\
         T-135 & ‘1000101001111110111111011' \\
         T-136 & ‘1000101001111110111111101' \\
         T-137 & ‘1000101001111110111111110' \\
         T-138 & ‘1000101001111110111111111' \\
         T-139 & ‘1000101001111111111111011' \\
         T-140 & ‘1000101001111111111111101' \\
         T-141 & ‘1000101001111111111111110' \\
         T-142 & ‘1000101001111111111111111' \\
         T-143 & ‘1000111101111111101010000' \\
         T-144 & ‘1000111101111111101010010' \\
         T-145 & ‘1000111101111111101010011' \\
         T-146 & ‘1000111101111111101010101' \\
         T-147 & ‘1000111111010100111111011' \\
         T-148 & ‘1000111111010100111111101' \\
         T-149 & ‘1000111111010100111111110' \\
         T-150 & ‘1000111111010100111111111' \\
         T-151 & ‘1000111111111111101010000' \\
         T-152 & ‘1000111111111111101010010' \\
         T-153 & ‘1000111111111111101010011' \\
         T-154 & ‘1000111111111111101010101' \\
         T-155 & ‘1001111101111110111111011' \\
         T-156 & ‘1001111101111110111111101' \\
         T-157 & ‘1001111101111110111111110' \\
         T-158 & ‘1001111101111110111111111' \\
         T-159 & ‘1001111101111111111111011' \\
         T-160 & ‘1001111101111111111111101' \\
         T-161 & ‘1001111101111111111111110' \\
         T-162 & ‘1001111101111111111111111' \\
         T-163 & ‘1001111111111110111111011' \\
         T-164 & ‘1001111111111110111111101' \\
         T-165 & ‘1001111111111110111111110' \\
         T-166 & ‘1001111111111110111111111' \\
         T-167 & ‘1001111111111111111111011' \\
         T-168 & ‘1001111111111111111111101' \\
         T-169 & ‘1001111111111111111111110' \\
         T-170 & ‘1001111111111111111111111' \\
         \vspace{2pt} 
    \end{tabular}
    \end{ruledtabular}
\end{minipage}
\end{table*}
\vspace{10pt}

\twocolumngrid
\section{Results for Quantum Trajectories Search}
\begin{figure*}[t!]
    \includegraphics[width=\textwidth]{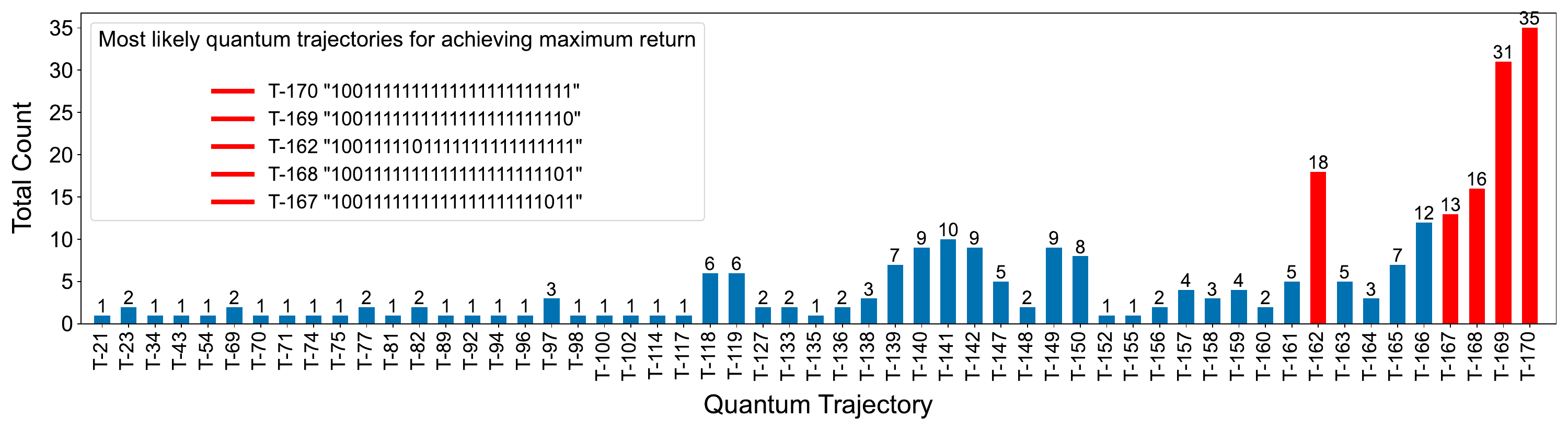}
    \caption{Distribution of quantum trajectories after executing Grover's algorithm to search the trajectories starting from any state and terminating at $s_3$ in the QMDP over 3 time steps. Each trajectory number represents unique trajectory sequences. The most likely high-return trajectories are highlighted in red. The trajectory ‘‘100111111111111111111111’’ was sampled 35 times, making it the most frequently identified optimal path for this scenario.}
    \label{fig:1001_Grover}
\end{figure*}

\vspace{-6pt}
\label{appendixB}
In this section, we examine Grover's search for identifying the optimal trajectories for broader search space where the agent can start from any initial state within the state space with equal probability and must terminate at state $s_3$, aiming to maximize the total return over 3 time steps. This scenario expands the search space by introducing uncertainty in the agent's starting state, making the task of finding the optimal trajectory more complex. The result of this demonstration, visualized in Fig.~\ref{fig:1001_Grover}, shows the distribution of quantum trajectories identified by Grover's search algorithm. The trajectory numbers on the x-axis represent distinct quantum trajectories, with each number representing a unique sequences of state-action-next state and reward over 3 time steps, followed by a return value (see Table~\ref{table:classical_trajectories} in Appendix \ref{appendixA} for the corresponding quantum trajectories). 
Based on the classical measurement of trajectory distribution in Fig.~\ref{fig:classical}, there are a total of 16 trajectories, from T-155 to T-170, that achieve the maximum return (‘1001’) under this scenario. Grover's search algorithm is capable of identifying these trajectories with a single call to the oracle. The results show that Grover's search efficiently determines these trajectories, and among them, the five trajectories with the highest frequencies are ‘‘100111111111111111111111’’, ‘‘100111111111111111111110’’, ‘‘100111101111111111111\\111’’, ‘‘100111111111111111111101’’, and ‘‘1001111111111\\11111111011’’. These represent the most likely optimal paths where the agent successfully terminates at state $s_3$ while achieving the maximum return. Although these five trajectories are the most frequently sampled under the given condition, other trajectories were observed as well, though with lower frequencies. These quantum trajectories indicate that action $a_1$ (‘1’) is the optimal action for all states in this scenario. 

\begin{figure}[h]
    \includegraphics[width=0.5\textwidth]{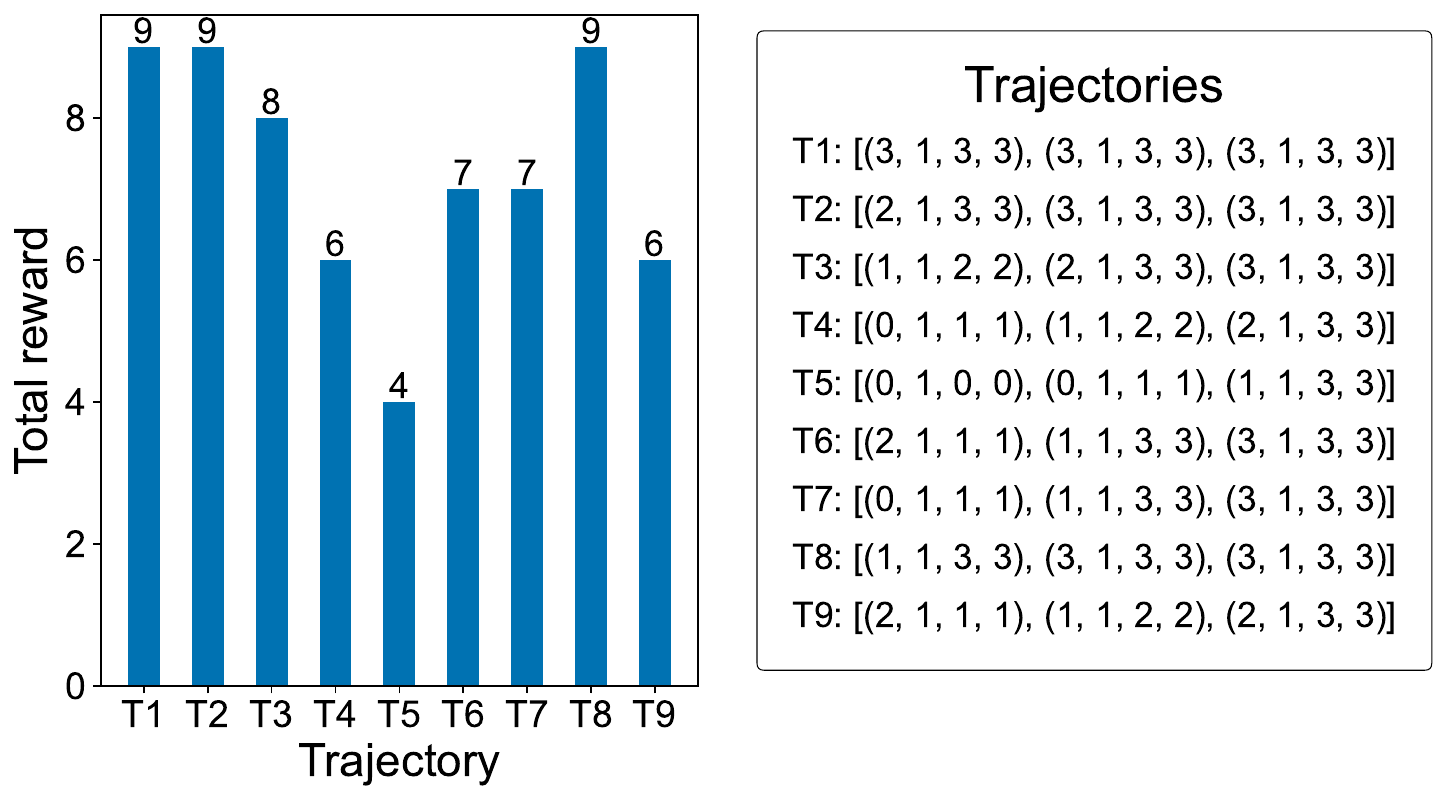}
    \caption{Comparison of total reward for each unique trajectory (T1–T9), starting from any state and ending at $s_3$ over 3 time steps in classical Q-learning.}
    \label{fig:reward_anystate}
\end{figure}

After Grover's search algorithm, we now compare the outcomes using classical Q-learning under the same scenario. The Q-table generated from the Q-learning process is shown in Table \ref{table:qvaluestartinganystates}. As highlighted in Table \ref{table:qvaluestartinganystates}, it reveals that action $a_1$ consistently provided the highest Q-value for all states. To identify optimal trajectories, we executed the Q-learning algorithm for 100 trials across 3 time steps. This process generated nine unique trajectories with total rewards ranging from 4 to 9, as illustrated in Fig.~\ref{fig:reward_anystate}. The bar chart illustrates the total rewards obtained for each unique trajectory that starts from any state and terminates at $s_3$ over 3 time steps. The x-axis lists each trajectory ($T1$ through $T9$), and the y-axis indicates the total reward associated with each trajectory. The bars are labeled with their respective reward values. Among these nine trajectories, the maximum reward ‘9’ can be achieved by these three trajectories ($T1, T2,$ and $T8$). In $T1$, the system consistently returns to state $s_3$ at each time step, maximizing cumulative reward. $T2$ starts in state $s_2$ but moves to and remains in $s_3$ for the subsequent time steps, while $T8$ makes an initial transition at state $s_1$ and eventually leads to state $s_3$, where it stays for the remaining time steps to achieve the highest total reward of 9.

\begin{figure}[h]
     \includegraphics[width=0.5\textwidth]{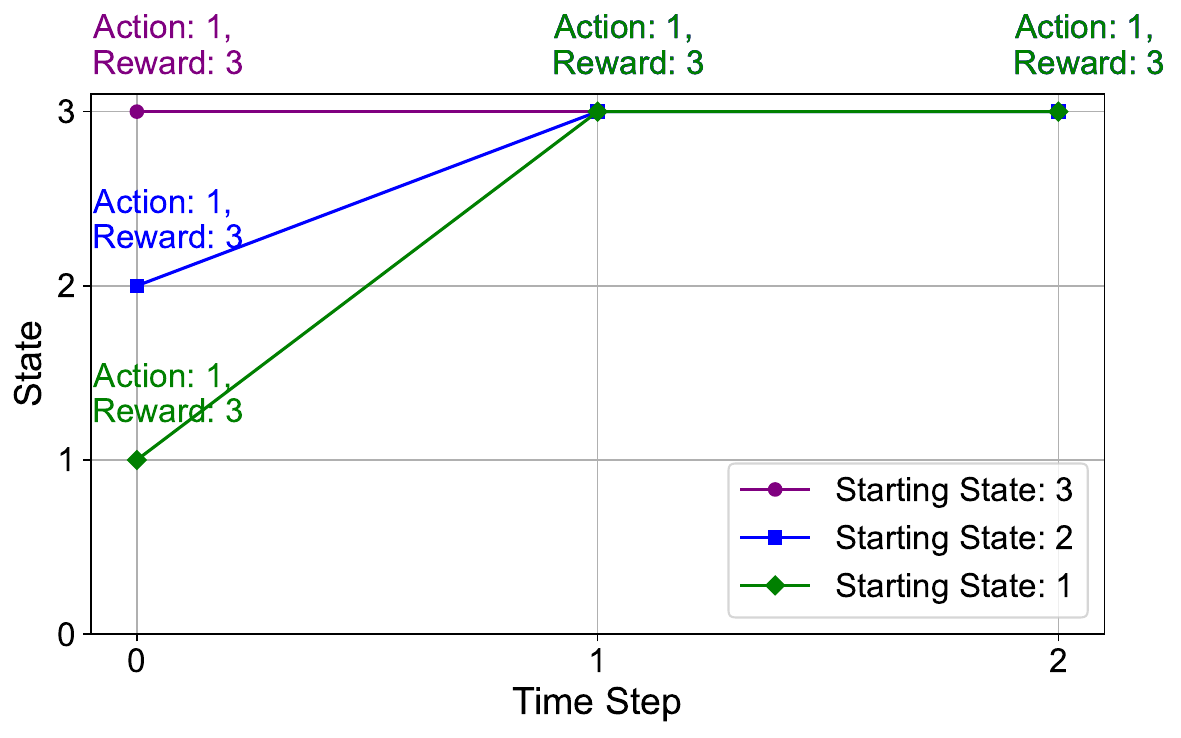}
    \caption{Optimal trajectories from 3 different starting states (1, 2, and 3) with each trajectory converging to the terminal state $s_3$. The x-axis represents the time steps, while the y-axis shows the state transitions. Each colored line represents a different starting state: purple for state 3, blue for state 2, and green for state 1.}
    \label{fig:optimal_anystate}
\end{figure}

Figure~\ref{fig:optimal_anystate} provides a detailed view of these optimal trajectories over 3 time steps, showing the agent's path from different starting states, and highlighting the actions and rewards at each time step. When examining the trajectories from Q-learning and Grove's search, $T1$ corresponds to the quantum trajectory ‘‘100111111111111111111111’’ which was identified with the highest count. Similarly, $T2$ and $T8$ are also among the top trajectories identified by Grover's search, represented as ‘‘100111111111111111111110’’ and ‘‘100111111111111111111101’’, respectively. These results demonstrate that Grover’s search is highly effective in identifying the most frequent quantum trajectories, even in the broader search space with no fixed initial state.
\begin{table}[h!]
\caption{\label{table:qvaluestartinganystates} Learned Q-values for state-action pairs for 4 states MDP, with the optimal action for each state highlighted in yellow. Action $a_1$ consistently yields the highest Q-value across all states ($s_0$ to $s_3$), indicating it as the optimal action choice in each state for maximizing cumulative reward.}
\begin{ruledtabular}
\centering
\renewcommand{\arraystretch}{0.7}  
\begin{tabular}{ccc}
 & $a_0$ & $a_1$  \\
  \hline
  \noalign{\vskip 3pt}
$s_0$ & 35.15 & \cellcolor{yellow} 36.67 \\
$s_1$ & 42.66 & \cellcolor{yellow} 46.87 \\
$s_2$ & 47.34 & \cellcolor{yellow} 48.73 \\
$s_3$ & 48.36 & \cellcolor{yellow} 49.20 
\end{tabular}
\end{ruledtabular}
\end{table}

\section{System and Simulation Environment Specifications}\vspace{-6pt}
\label{appendixC}
The computations in this work were conducted using the IBM Qiskit framework to simulate quantum circuits on a classical device. Below, we provide detailed specifications of both the quantum simulation environment and the classical hardware used in this study.

Quantum simulation environment specifications:
\begin{enumerate}
[noitemsep]
  \item Quantum framework: IBM Qiskit (version 1.1.0)
  \item Quantum simulator: IBM Qiskit Aer Simulator (version 0.14.2)
  \item Simulation method: Statevector simulation
  \item Simulation device: CPU-based simulation
  \item Noise model: None (ideal simulation)
\end{enumerate}

Classical device specifications:
\begin{enumerate}
[noitemsep]
  \item Processor: AMD Ryzen 5 PRO 5650U with Radeon Graphics, 2.30 GHz
  \item RAM: 16 GB
  \item System type: 64-bit operating system, x64-based processor
\end{enumerate}
\twocolumngrid
\bibliography{reference_list}

\end{document}